\documentclass[%
 reprint,
 amsmath,amssymb,
 aps,
 prl
prb]{revtex4-2}
\usepackage{physics}
\usepackage{graphicx}
\usepackage{dcolumn}
\usepackage{xcolor} 
\usepackage{chngcntr}

\usepackage{bm}
\usepackage[normalem]{ulem} 

\usepackage[colorlinks=true,
            linkcolor=blue,
            citecolor=blue,
            urlcolor=blue]{hyperref}
\usepackage{subfiles}

\newcommand{\kistrike}[1]{%
  {\color{blue}%
    \ifmmode
      \text{\sout{\(\mathstrut #1\)}}%
    \else
      \sout{#1}%
    \fi
  }%
}

\newcommand{\sq}[1]{\left[#1\right]}

\newcommand{\rnd}[1]{\left(#1\right)}

\newcommand{\bld}[1]{\boldsymbol{#1}}

\newif\ifprl
\prlfalse 

\newcommand{\suppRef}[2]{%
  \ifprl
    \cite{SM_Key}
  \else
    #1 Appendix~\ref{#2}
  \fi
}

\begin{document}


\title{Dispersion of Anyon Bloch Bands}

\author{Kishore Iyer$^1$}
\author{Andreas Feuerpfeil$^2$}
\author{Valentin Cr\'epel$^3$}
\author{Nicolas Regnault$^{4, 5, 6}$}
\author{Christophe Mora$^1$}

\vspace{5pt}
\affiliation{$^1$Universit\'e Paris Cit\'e, CNRS, Laboratoire Mat\'eriaux et Ph\'enomènes Quantiques, 75013 Paris, France}
\affiliation{$^2$Institut f\"ur Theoretische Physik und Astrophysik and W\"urzburg-Dresden Cluster of Excellence ct.qmat, Julius-Maximilians-Universit\"at W\"urzburg, Am Hubland, Campus S\"ud, W\"urzburg 97074, Germany}
\affiliation{$^3$Department of Physics, University of Toronto, 60 St.~George Street, Toronto, ON, M5S 1A7 Canada}

\affiliation{$^4$Center for Computational Quantum Physics, Flatiron Institute, 162 5th Avenue, New York, NY 10010, USA}
\affiliation{$^5$Department of Physics, Princeton University, Princeton, New Jersey 08544, USA}
\affiliation{$^6$Laboratoire de Physique de l'Ecole normale sup\'{e}rieure, ENS, Universit\'{e} PSL, CNRS, Sorbonne Universit\'{e}, Universit\'{e} Paris-Diderot, Sorbonne Paris Cit\'{e}, 75005 Paris, France}

\date{\today}

\begin{abstract}
Fractional Chern insulators (FCIs) are zero magnetic field analogs of  fractional quantum Hall states. 
While the electrons forming an FCI are not subject to an external magnetic field, their anyonic excitations experience a  magnetic field with finite-flux due to a many-body Berry phase,  whose lattice periodicity generically induces some dispersion. 
From Laughlin wavefunctions at filling $\nu =1/m$, we analytically construct single-anyon Bloch states in an ideal band,  providing a basis to efficiently compute the dispersion. The anyon spectrum exhibits an $m$-fold degeneracy in the reduced magnetic Brillouin zone (BZ), which originates from the topological degeneracy of the FCI. From our wavefunctions, we derive the $m^2-$fold degeneracy seen in previous works, showing it to be a splicing of anyon momenta into the electronic BZ. Finally, we find that the anyon dispersion bandwidth is controlled by quantum geometry non-uniformity, growing linearly at weak modulation and saturating at strong modulation. Remarkably, higher harmonics of the quantum geometry alone strongly suppress the dispersion, which we attribute to emergent magnetic translation symmetries. When combined with the first harmonic, a positive (negative) second harmonic drives the system toward a second- (first-) harmonic-dominated regime, thereby reducing (enhancing) the bandwidth.
Our results offer an analytically controlled method for evaluating anyon spectra in ideal band FCI, shedding light on how non-uniform quantum geometry and emergent symmetries shape the dispersion of anyons.
\end{abstract}

\maketitle

\emph{Introduction---}Fractional quantum Hall (FQH) phases~\cite{stormer1999a} are the paradigmatic topologically ordered states of matter, hosting anyons, quasi-particle excitations~\cite{laughlin1983a, halperin1984statistics, jain1989a} that obey statistics intermediate between those of fermions and bosons~\cite{Leinaas1977, wilczek82, arovas84}. 
Owing in part to their promise for technological applications~\cite{nayak08, kitaev2003fault}, the past decades have witnessed an intense experimental pursuit of anyons in FQH systems, culminating in resounding successes in recent years~\cite{bartolomei2020a, banerjee2018observation, nakamura2020a, lee2023a, kundu2023a, ruelle2023a,iyer24, glidic2023a, nakamura2023a,  rosenow2025braidsbeamsexploringfractional, Feldman2021-fj}. 
Strong magnetic fields in FQH systems impose continuous magnetic translation symmetry (CMTS), which severely constrains the dynamics of charged (quasi-)particles and renders isolated anyons effectively immobile. 
While this pinning is responsible for the stability of FQH states under anyon doping, reflected in extended conductance plateaus, it impedes the realization and observation of itinerant anyonic phases~\cite{Shi2025}.

Fractional Chern insulators (FCIs)~\cite{kol1993fractional,sorensen2005fractional,aguado2008creation,moller2009composite,goldman2009non,kapit2010exact,neupert2011a, sun2011a, sheng2011a, tang2011a, wang2011a, regnault2011a, bergholtz2013a, parameswaran2013a} realize physics analogous to that of FQH systems -- including fractionally quantized Hall conductivity and anyonic excitations -- in the absence of magnetic field, relying instead on its momentum-space analog: the Berry curvature~\cite{thouless1982a, haldane1988a}. 
Their recent experimental observation in twisted $\text{MoTe}_2$ (t$\text{MoTe}_2$)~\cite{cai2023a, park2023a, xu2023a, zeng2023a, ji2024local, redekop2024direct} and rhombohedral pentalayer graphene~\cite{lu2024a, xie2025tunable, lu2025extended} represents a new frontier in the physics of anyonic excitations~\cite{zhang2019a, crepel2023anomalous, ledwith2020a, crepel2019matrix, repellin2020a, wilhelm2021a, nakatsuji2025high, abouelkomsan2020a, wu2019a}. 
While anyons in FCIs share the same topological properties as their FQH counterparts, the presence of an underlying lattice fundamentally alters their dynamics by breaking continuous (magnetic) translation symmetry down to a discrete one. As a result, doping an FCI with anyons can give rise to emergent itinerant anyonic phases~\cite{Shi2025}, including anyon superconductivity~\cite{laughlin1988a, fetter1989a, lee1989a, chen1989a, halperin1989a, divic2025a, Pichler_2025, zhang2025holonmetalchargedensitywavechiral, kuhlenkamp2025robust, wang2025chiralsuperconductivitynearfractional}. 
Superconducting phases might have been experimentally observed in the vicinity of FCI states in $\text{tMoTe}_2$~\cite{xu2025signatures}, stimulating rich theoretical developments~\cite{shi2025anyondelocalizationtransitionsdisordered, nosov2025anyonsuperconductivityplateautransitions, crepel2024attractive}.

These advances have motivated detailed investigations of anyon dispersions in FCI systems. 
General  field-theoretic arguments for topologically ordered phases~\cite{barkeshli2009structure, cheng2016a, bultinck2018a, Shi2025} predict an $m$-fold degeneracy for quasihole excitations of charge $e^* = -e/m$ in abelian Laughlin-like states.
In contrast, recent exact diagonalization (ED) studies of realistic model of twisted $\text{tMoTe}_2$~\cite{goncalves2025} have characterized the non-trivial dispersion of anyons and obtained
an intriguing $m^2$-fold degeneracy at $\nu = 2/3$~\cite{xiaodong2024, liu2025}. 
Related ED of FQH systems subject to non-uniform magnetic fields on the sphere confirmed that interactions and non-uniform quantum geometry yields non-trivial anyon dispersion~\cite{schleith2025}.

Trial wavefunctions offer a complementary approach, allowing analytical control over the physics of FQHE states~\cite{laughlin1983a, jain1989a, moore1991a}. Not only do they faithfully capture universal low-energy topological order, but they also reproduce microscopic details, showing quantitative agreement with ED, including for bulk and edge anyonic excitations~\cite{jain1989a,yang2012model,balram2013state,crepel2019microscopic}.
Furthermore, trial wavefunctions enable investigating larger system sizes through Monte Carlo (MC) methods~\cite{wang19, wang2026hybridmontecarlofractional}.

A similar route can be taken to describe special cases of FCI hosted in ideal bands~\cite{roy2014, Ledwith2023, Wang2021, Estienne2023, crepel2023chiral, meraozawa2024, crepel2025topologically}. Ideal bands can be expressed as a product of a normalized spinor $\chi(\bld{r})$ and a lowest Landau level (LLL) wavefunction in a periodic magnetic field characterized by a Poisson field $ \mathcal{B}(\bm{r})\,,$ closely mimicking the LLL in that attaching vortices keeps the wavefunction within the same band. 
This property allows construction of exact many-body FCI trial wavefunctions,  which provide an ideal platform for analytical and numerically scalable investigations of anyon dispersions in FCIs~\cite{yan2025, liu2025}.

In this work, we start from Laughlin-like FCI single-anyon (quasihole) wavefunctions on a torus~\cite{haldane1985a, Greiter2016, Pu2017}. Due to non-uniform quantum geometry, the anyon experiences a position-dependent potential. We show that while electrons in an FCI are non-magnetic, obeying ordinary crystalline translation symmetries~\cite{Guerci2025}, anyons are magnetic due to an emergent many-body Berry phase. Hence, anyon translations in the two directions do not commute, rendering them conjugate variables. Consequently, even in a flat electronic band~\cite{tarnopolsky19}, anyons acquire a dispersion from the interplay of interactions and quantum geometry.

Our objective is to compute the dispersion of single-anyon states, understand its fundamental properties and its connection to the universal topological properties of the parent FCI, and analyze how the dispersion evolves with quantum geometry modulations. To this end, we construct anyon Bloch wavefunctions, eigenstates of the anyon translation operators. We compute the energies of anyon Bloch states using both MC and ED, considering the Coulomb interaction. Even when the single-particle dispersion is flat, we find a non-trivial anyon dispersion with an $m$-fold periodicity in the reduced Brillouin zone (BZ) of anyons consistent with effective field theory arguments~\cite{cheng2016a, bultinck2018a, Shi2025}. We explicitly show the origin of the $m$-fold periodicity to lie in the topological degeneracy of the FCI state on the torus. 
Previous ED studies~\cite{xiaodong2024,goncalves2025, liu2025} reported an $m^2$-fold periodicity of the anyon dispersion which was explained with heuristic arguments~\cite{goncalves2025, yan2025} but a \emph{direct proof} was lacking. Here, we show at the level of the many-body wavefunction  that the $m^2$-fold structure arises from a redundant splicing of the $m$-fold spectrum as a function of the anyon crystal momentum into the electronic BZ, which we establish using an analytical relation between anyon and electron translations. 

We further investigate how the anyon dispersion evolves with the strength and structure of quantum geometry non-uniformity. We find that the bandwidth grows linearly at weak modulation and saturates at strong modulation. Surprisingly, higher harmonics strongly suppress the dispersion: the second harmonic alone produces an almost flat anyon dispersion. We trace this suppression to enhanced  exact symmetries, indicating that higher harmonics drive the system back toward a uniform quantum geometry.

\emph{FCI in ideal bands---}The single anyon (quasihole) FCI ideal band wavefunction on a torus  takes the form $\psi^\sigma[\{\boldsymbol{r}_i\}, \boldsymbol{\eta}] = \psi_L^\sigma[\{\boldsymbol{r}_i\}, \boldsymbol{\eta}]\prod_{i=1}^{N_e}\chi_i(\bld{r}_i)$~\cite{Greiter2016, ledwith2020a}, with:
\begin{equation}
\begin{split}
    \psi_L^\sigma[&\{\boldsymbol{r}_i\}, \boldsymbol{\eta}] = A_\sigma e^{\frac{1}{4ml^2}\mathcal{B}(\boldsymbol{\eta})} e^{iK_\sigma\rnd{Z + \frac{\eta}{m}}} 
    \prod_{i<k}^{N_e}\vartheta_1\sq{\frac{z_i-z_k}{L_1}\Bigg| \tau}^m \\
    &\prod_{\gamma = 1}^m \vartheta_1\sq{\frac{Z + \frac{\eta}{m} - Z^\gamma_\sigma}{L_1} \Bigg| \tau}  \prod_{i = 1}^{N_e} \vartheta_1\sq{\frac{z_i - \eta}{L_1}\big| \tau} e^{\frac{1}{4l^2}\mathcal{B}(\boldsymbol{r_i})}\,, 
\end{split}
\label{eq:torus_wavefunction}
\end{equation}
where $\psi_L$ is a Laughlin-quasihole wavefunction on the torus under a non-uniform magnetic field, and $A_\sigma$ is a normalization factor~\suppRef{with the details given in}{app:wavefunction}. All coordinates above are two-dimensional; the bolded coordinates are vectors $\boldsymbol{r} = (x, y)$, while the unbolded ones are in complex notation $z = x+ iy$. The wavefunction is defined on a $2$D torus spanned by complex vectors $L_j$ with $j = 1,2$ and $\tau = L_2/L_1$, hosting $N_e$ electrons \footnote{The term electrons is used to generically denote the microscopic constituents; for $m$ even, these are bosons.} at positions $z_i$ and one anyon at position $\eta$, with $Z = \sum_i z_i$. The magnetic length is given by $l^2 = \text{Im}(\bar{L}_1L_2)/(2\pi N_\phi)$, where $N_\phi = mN_e +1$ is the number of flux quanta, and $m \in \mathbb{Z}$ is the inverse Laughlin filling fraction. 
We have $N_j$ unit cells in the $j$ direction, defining $l_j = L_j/N_j$. Taking $N_1N_2 = N_\phi$, each unit cell encloses a single flux quantum by construction. Finally, $\vartheta_1(z|\tau)$ denotes the elliptic theta function and $K_\sigma, Z^\gamma_\sigma$ are complex numbers that encode the topological degeneracy of the FCI on a torus~\cite{Greiter2016}, indexed by $\sigma \in \{1,\dots,m\}$.

\emph{Electron and anyon translations---}$\mathcal{B}(\boldsymbol{r})$ encodes the quantum geometry of the single particle Chern band ~\cite{meraozawa2021,meraozawa2021b, meraozawa2021c}. The bands carry a unit flux per unit cell that we parametrize  as $\mathcal{B}(\boldsymbol{r}) = z^2 -|z|^2 - \mathcal{K}(\boldsymbol{r})$. The first two terms account for a uniform unit magnetic flux per unit cell, 
while $\mathcal{K}(\boldsymbol{r}+\boldsymbol{l_j}) = \mathcal{K}(\boldsymbol{r})$ is non-uniform, averaging to zero over a unit cell. A non-uniform $\mathcal{K}(\boldsymbol{r})$ coupled with a layer skyrmion in $\chi(\boldsymbol{r})$ breaks the CMTS of electrons down to  discrete translation symmetry over a unit cell in FCI.  Indeed, under electron translations over a unit cell,  the Laughlin-quasihole wavefunction picks up a Berry phase $e^{\frac{i}{2l^2}\boldsymbol{l}_j\times\boldsymbol{r}}$, coming from the term $e^{\frac{1}{4l^2}\mathcal{B}(\boldsymbol{r_i})}$.
Meanwhile, the spinor $\chi(\boldsymbol{r})$ possesses an orbital layer skyrmion with a Berry phase such that $\chi(\boldsymbol{r}+\boldsymbol{l_j}) = e^{-\frac{i}{2l^2}\boldsymbol{l}_j\times\boldsymbol{r}}\chi(\boldsymbol{r})$. Hence, the Berry phases coming from the K\"ahler potential and the normalized spinor cancel each other out~\cite{Guerci2025}. Consequently, electrons in
FCI locally feel a non-uniform magnetic field, but no net magnetic field.

Surprisingly, unlike electrons, anyons in a zero-field FCI do feel a net non-zero magnetic field. Under lattice translations of the anyon, the wavefunction picks up a Berry phase $e^{\frac{i}{2ml^2}\boldsymbol{l}_j\times\boldsymbol{r}}$. In contrast to the electronic case, there is no spinor term that cancels this Berry phase. 
As a result, even in the absence of an external magnetic field, the many-body Berry phase of the FCI wavefunctions renders anyons magnetic: they experience a non-uniform magnetic field of $1/m$ flux quanta per unit cell, under which the position operators in the two spatial directions no longer commute and instead act as conjugate variables. Consequently, even when the single-particle (electronic) dispersion is flat, interactions together with non-uniform quantum geometry can generate a nontrivial dispersion for the anyons.

We now define anyon \emph{magnetic} translation operators over a unit cell (aMTO)
$T_1, T_2$ that act on Eq.~\eqref{eq:torus_wavefunction} by moving the anyons and multiplying the wavefunction by a compensating phase factor \suppRef{as shown in}{app:translations}. The aMTO satisfy projective commutation relations~\cite{cheng2016a, bultinck2018a, Shi2025}: 
\begin{equation}
    T_1 T_2 = e^{2\pi i/m}T_2 T_1\,.
\end{equation}
We further define the operators $\mathcal{T}_j \equiv T_j^{N_j}$, that move the anyon over the entire system in either direction.  The action of $\mathcal{T}_2$ on the wavefunction in Eq.~\eqref{eq:torus_wavefunction} is to shift the topological sector \cite{Greiter2016}, $\mathcal{T}_2 \psi^\sigma = \psi^{\sigma+1}$ \suppRef{as detailed in}{app:topological_degeneracy}.

\emph{Anyon Bloch functions---}The aMTO, $T_1$ and $T_2^m$ commute with each other, defining an $m-$fold enlarged unit cell. Since $\mathcal{K}(\boldsymbol{\eta}+\boldsymbol{l_j}) = \mathcal{K}(\boldsymbol{\eta})$, the anyon feels the same background potential when translated by multiples of unit cells. Thus, $T_1, T_2^m$ also commute with the Hamiltonian projected to the subspace of anyon Bloch wavefunctions,
forming a convenient basis to label energy eigenstates and deduce spectral properties~\cite{Bernevig2012, Shi2025}. This motivates us to construct Bloch-like wavefunctions that take the form:
\begin{equation}
    \Psi_{k_1,k_2}[\{\boldsymbol{r}_i\}] = \frac{1}{\mathcal{N}(\bld{\eta})}\sum_{\boldsymbol{n}}e^{-2\pi i \boldsymbol{k}\cdot \boldsymbol{n} }~T_1^{n_1}T_2^{mn_2}\psi^\sigma[\{\boldsymbol{r}_i\}, \boldsymbol{\eta}]\,,
    \label{eq:anyon_bloch_function}
\end{equation}
where $\mathcal{N}(\bld{\eta})$ is a normalization factor, $\boldsymbol{k} = (k_1, k_2)$, $\boldsymbol{n} = (n_1, mn_2 )$, with $n_{1/2} \in \{0,\dots N_{1/2}-1\}$. The momentum quantum numbers $k_1, k_2$ span the anyon magnetic BZ such that $k_1 \in [0, 1)$, while $k_2 \in [0, 1/m)$. The anyon Bloch wavefunctions satisfy the relations:
\begin{equation}
\begin{split}
    T_1  \Psi_{k_1,k_2}[\{\boldsymbol{r}_i\}\,, \boldsymbol{\eta}] &= e^{2\pi i k_1 } \Psi_{k_1,k_2}[\{\boldsymbol{r}_i\}\,, \boldsymbol{\eta}]\,, \\
    T_2^m \Psi_{k_1,k_2}[\{\boldsymbol{r}_i\}\,, \boldsymbol{\eta}] &= e^{2\pi i m k_2} \Psi_{k_1,k_2}[\{\boldsymbol{r}_i\}\,, \boldsymbol{\eta}]\,,
\end{split}   
\label{eq:bloch_relations}
\end{equation}
provided the momenta $k_{1/2}$ satisfy the conditions:
\begin{equation}
\begin{split}
    e^{2\pi i k_1 N_1}  &= e^{\frac{2\pi i}{m} (\sigma-1) }\,, \\
    e^{2\pi i m k_2 N_2} &= 1\,,
\end{split}
\label{eq:momenta_constraint}
\end{equation}
where we emphasize that $\sigma$ indicates the topological index of the wavefunction in Eq.~\eqref{eq:anyon_bloch_function}\suppRef{; the details are laid out in}{app:commuting_translations}. In a finite-size system, there are exactly $N_\phi$ allowed momenta, taking values on a grid dictated by Eq.~\eqref{eq:momenta_constraint}. Crucially, the grid depends on the topological sector: since $N_1N_2=mN_e+1$ and thus $\text{gcd}(m, N_j) = 1$, the momenta allowed in the different topological sectors are fully different from each other.

\emph{$m$-fold Degeneracy of Anyon Bloch Bands---}The action of the operator $T_2$ on the anyon Bloch functions reveals an interesting $m-$fold degeneracy~\cite{cheng2016a, bultinck2018a, Shi2025}. We find\suppRef{, as shown in}{app:T2}:
\begin{equation}
    T_2 \Psi_{k_1,k_2}[\{\boldsymbol{r}_i\}] = \Psi_{k_1+1/m\,,~k_2}[\{\boldsymbol{r}_i\}]\,.
\end{equation}
Since $k_1$ is defined modulo $1$, the action of $T_2$ induces a cyclic shift of $k_1$ by a value $1/m$ within an $m-$dimensional subspace. Combined with Eq.~\eqref{eq:bloch_relations}, this defines a clock algebra within each $m-$dimensional subspace of the Bloch bands. Within the latter subspace, the operators $T_1$ and $T_2$ can be represented as:
\begin{equation}
\rnd{T_1}_{ij} = e^{2\pi i k_1} \omega^{j-1} \delta_{ij}\,, \qquad
\rnd{T_2}_{ij} = \delta_{i, (j+1)~\text{mod}~m}
\end{equation}
where $\omega = e^{2\pi i/m}$, implying an $m-$fold degeneracy (in the $k_1$ eigenvalues) within the anyon magnetic BZ. 

This degeneracy is in fact directly related to the $m-$fold topological degeneracy of the Laughlin-like FCI wavefunction on a torus. The momenta $k_1 + 1/m$ are incommensurate with  Eq.~\eqref{eq:momenta_constraint}, for a fixed topological sector $\sigma$, but they are compatible with the topological sector $\sigma + q_1$ with $q_{1/2} = N_{2/1} ~\text{mod}~m$, connecting the $m$ topologically degenerate states.  Indeed, when acting on the Bloch wavefunctions, $T_2 =[e^{2\pi i m k_2}]^{r} \mathcal{T}_2^{q_1}$, where $r$ is an integer \suppRef{as shown in}{app:T2_topological_degeneracy}.
This confirms that the $m$-fold degeneracy of the anyon Bloch states originates from the topological degeneracy of the FCI wavefunctions, as predicted by earlier effective field theory studies.
To illustrate the analytical derivation, we numerically compute the Coulomb energies of the anyon Bloch wavefunctions 
\begin{equation}
    E_{\boldsymbol{k}} = \frac{\int \prod_{i=1}^{N_e} d^2z_i ~|\Psi_{k_1,k_2}[\{\boldsymbol{r}_i\}]|^2 ~V(\{\boldsymbol{r}_i\})}{\int \prod_{i=1}^{N_e} d^2z_i ~|\Psi_{k_1,k_2}[\{\boldsymbol{r}_i\}]|^2}
    \label{eq:energy_defn}
\end{equation}
using the Metropolis-Hastings algorithm, where $V(\{\boldsymbol{r}_i\})$ is the pairwise Coulomb potential on the torus~\cite{Pu2017}. For the sake of illustration, we take a simple K\"ahler potential 
\begin{equation}
 \mathcal{K}(\boldsymbol{r}_i) = \sum_{n=1,2}4s_{n} \sum_{c=1,2}\frac{1}{|n\vec{b}_c|^2} \cos(n\vec{b}_c\cdot \vec{r}_i) 
\end{equation}
where $s_n$ is the strength of the $n$th harmonic, and $\vec{b}_{1/2}$ are reciprocal lattice vectors of the torus. For simplicity, all our calculations are performed on a torus with square aspect ratio. For layer-independent potentials, Eqs.~\eqref{eq:torus_wavefunction} and~\eqref{eq:energy_defn} imply that the energy is independent of the normalized spinor.

\begin{figure}
    \centering
    \includegraphics[width=\linewidth]{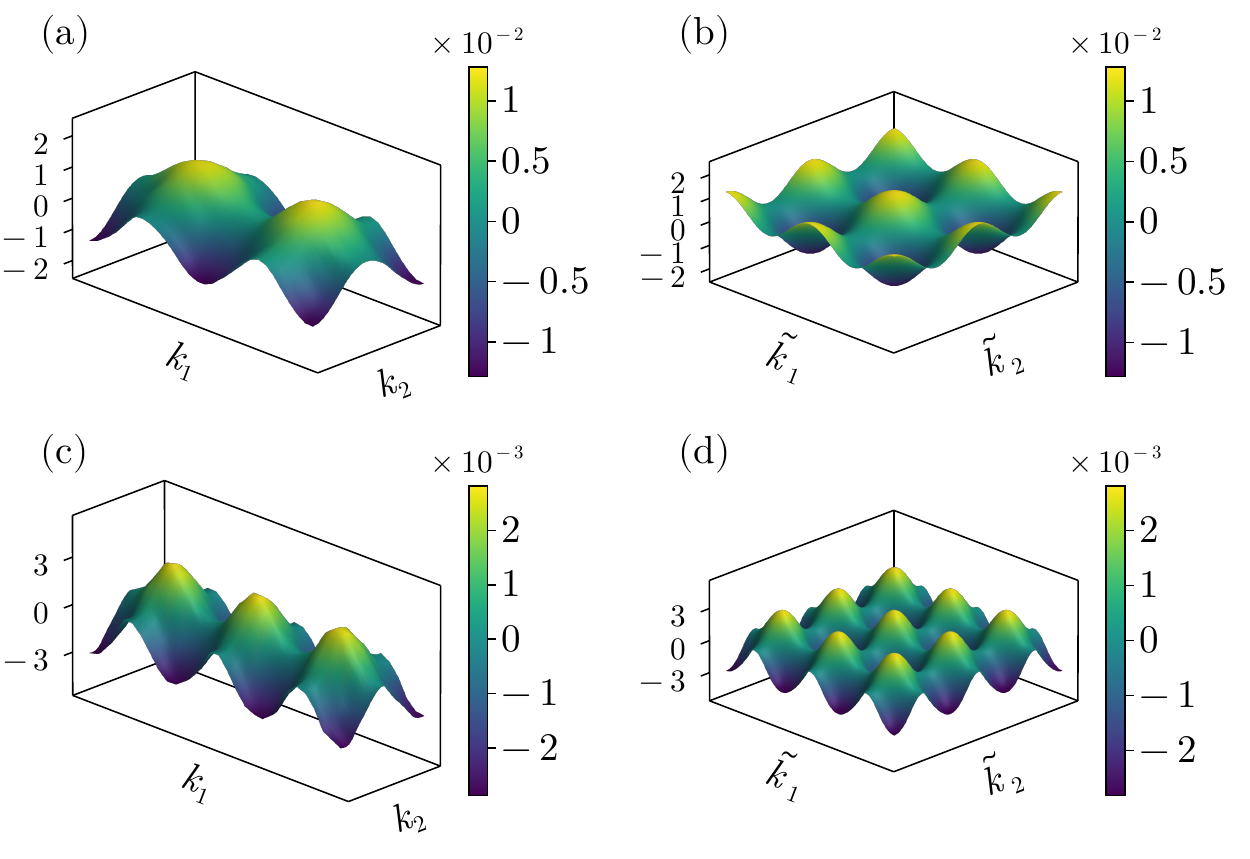}

    \caption{\textbf{Anyon dispersion spectra.} Coulomb interaction energy of anyon Bloch states on a square torus for the bosonic system $m=2$ [(a)-(b)], and the fermionic system $m = 3$ [(c)-(d)], both at system size $N_\phi = 25$  and $s_1 = 1,~s_2=0$. Results in (a)  \& (c) are obtained from MC simulations, while those in (b)  \& (d) are from ED on anyon Bloch wavefunctions. The energies  are shown in units of $e^2/(\epsilon l)$. In (a)  \& (c), the energies are plotted with respect to anyon momenta, showing an $m-$fold periodicity in the anyon magnetic BZ, as predicted by analytical arguments. In (b)  \& (d),  the energies are shown with respect to electron centre of mass momenta in the electronic BZ. The spectrum displays an $m^2-$fold periodicity in this case, consistent with our analytical relation between the anyon and electronic momenta.}
    \label{fig:anyon_momenta_mfold}
\end{figure}

The energies are shown as function of the anyon momenta in the anyon magnetic BZ in Figure~\ref{fig:anyon_momenta_mfold} for two systems: $m = 2$  (Figure~\ref{fig:anyon_momenta_mfold}(a)), and $m = 3$ (Figure~\ref{fig:anyon_momenta_mfold}(c)) both at system size $N_\phi = 25$ and $s_1 = 1,~ s_2 = 0$. The degeneracy is not exact in finite-size systems since an $m-$fold degeneracy is incompatible with $\gcd(N_{1/2},m) = 1$. However, the thermodynamic limit can be mimicked by flux insertion combined with a smooth interpolation of the finite-size data points, where the $m-$fold degeneracy in the $k_1$ momenta becomes manifest. 

\emph{$m^2-$fold Degeneracy of Anyon Bloch Bands---}Analytical arguments imply an $m-$fold degenerate energy spectrum of the anyon Bloch bands. Nevertheless, earlier exact diagonalization studies have observed $m^2-$fold degenerate spectra -- when energies are labelled by electron center of mass (CoM) momenta~\cite{goncalves2025, liu2025}. Ref.~\cite{goncalves2025} justified the observation by a heuristic analogy to the Hofstadter problem, but a \emph{direct proof} was lacking. We now provide such a proof from microscopic wavefunctions that the latter observation stems from topology.

We first define the electron CoM unit translations, as an operation that translates all the electrons in the system by one unit cell, denoted $\tilde{T}_j$. The operators $\tilde{T}_1, \tilde{T}_2$ commute with each other \suppRef{as shown in}{app:translations} and the Hamiltonian, forming an alternative basis to label energy eigenstates. The aMTO and the electron CoM unit translation operators obey the following relation \suppRef{derived in}{app:electron_quasihole_translation_relation}:
\begin{equation}
    \tilde{T}_{j} = T_{j}^{q_jN_j-1}\,,
\end{equation}
where $q_{1/2} = N_{2/1} ~\text{mod}~m$, and $j = 1, 2$ are the two spatial directions. This relation is physically intuitive; it implies that translating all the electrons in one direction is equivalent to translating the anyon in the opposite direction (up to a change in the topological sector for $j = 2$).

It follows that the electron CoM momenta $\tilde{k}_j$, defined in the domain $[0,  1)$, are related to the anyon momenta by:
\begin{equation}
    \tilde{k}_{j} = [k_{j}(q_{j}N_{j}-1)]~\text{mod}~1
    \label{eq:electron_anyon_momenta_relation}
\end{equation}
up to a constant shift. For $j = 1$, this operation is a rigid isometry, preserving the $m-$fold periodicity in the $\tilde{k}_1$ momenta \suppRef{as expained in}{app:electron_quasihole_translation_relation}. For $j=2$ on the other hand, the modulus redistributes consecutive momenta in anyon magnetic BZ spanning $[0, 1/m)$ into one of the $m$ domains $[0, 1/m),\dots, [1-1/m ,1)$ of the electronic BZ. In the thermodynamic limit, the $k_2$ values infinitesimally close to each other carry an almost identical energy. Distributed across the $m$-domains in the electronic BZ, this leads to an $m-$fold increase of degeneracy in the $k_2$ eigenvalues.  This is consistent with ED on anyon Bloch wavefunctions, as shown in Figures~\ref{fig:anyon_momenta_mfold}(b) and (d), and with Ref. \cite{yan2025}. The anyon Bloch wavefunctions are obtained in ED as zero modes of Haldane pseudopotentials, and the energy is computed as the expectation value of  Coulomb interaction in these states. To increase the resolution of the figures, we perform flux insertion \suppRef{as detailed in}{sup:ED}. While the $m-$fold degeneracy in $\tilde{k}_1$  has the same topological origin, the one in $\tilde{k}_2$ is a copying of close states in the anyon magnetic BZ into different domains of the electronic BZ.

\begin{figure}
    \centering
    \includegraphics[width=\linewidth]{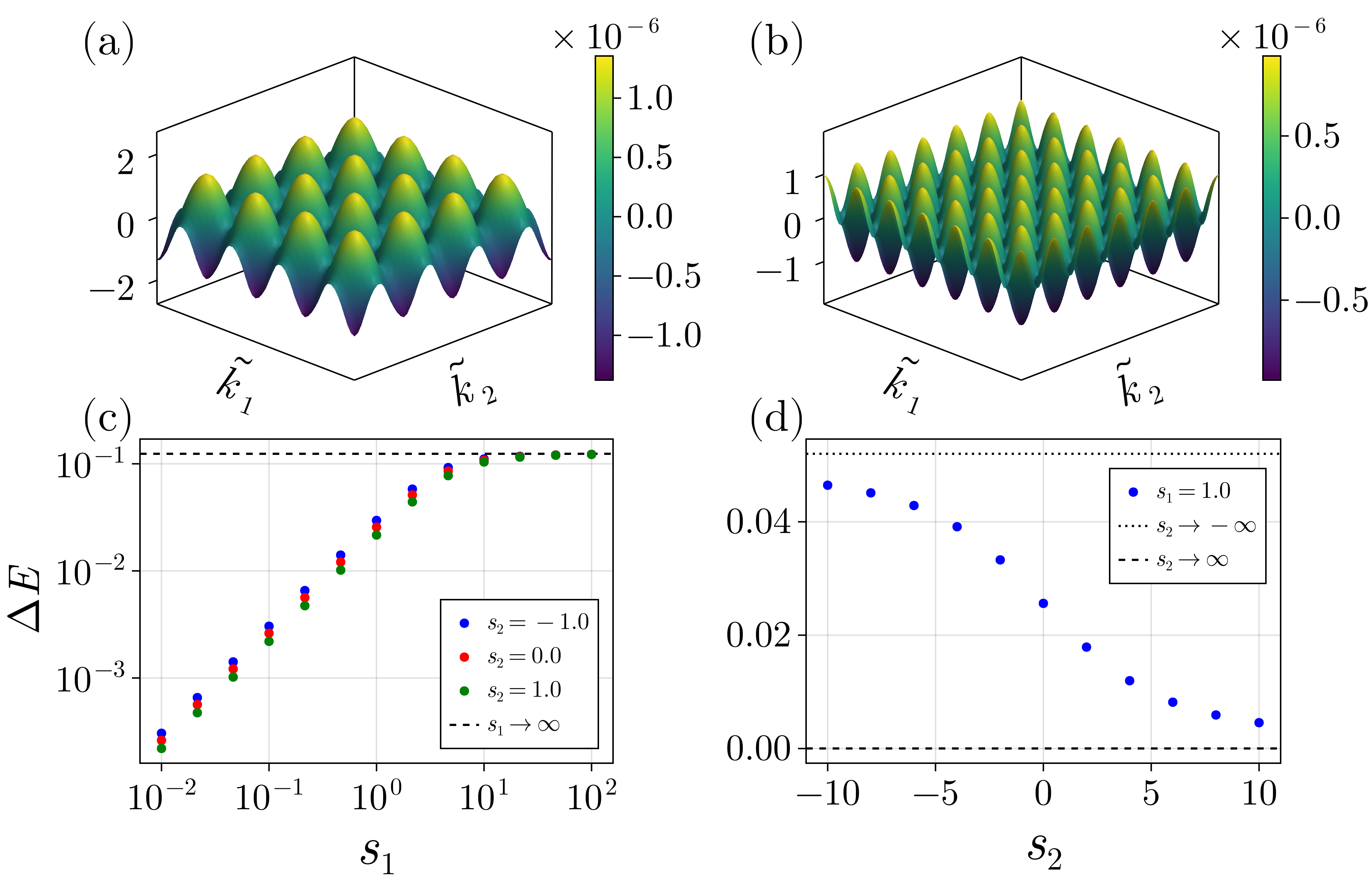}

    \caption{\textbf{Higher harmonics and bandwidth.} (a)-(b) Anyon dispersion at $s_1=0, ~s_2=1$ for (a) $m=2$ and (b) $m=3$ both at $N_\phi=25$. The dispersion exhibits a $(2m)^2$-fold degeneracy. (c)-(d) Bandwidth ($\Delta E$) evolution with quantum geometry for $m=2$, $N_\phi=25$. (c) $\Delta E$ as a function of $s_1$ at fixed $s_2$. On a square torus, $\Delta E$ is symmetric under $s_1 \rightarrow -s_1$ \suppRef{as detailed in}{app:symmetries_rectangular}; we only show the $s_1 >0$ part here. $\Delta E$ scales linearly for small $s_1$. While positive $s_2$ suppresses $\Delta E$, negative $s_2$ enhances it. In the large $s_1$ limit, all curves converge to a common saturation value as the first harmonic dominates the localization physics.
 (d) $\Delta E$ vs. $s_2$ at fixed $s_1$. Increasing $s_2$ suppresses $\Delta E$ toward a saturated minimum, effectively restoring band flatness. Conversely, decreasing $s_2$ drives an increase in $\Delta E$ as it reinforces the $s_1$ minima. }     
\label{fig:bandwidth}
\end{figure}

\emph{Evolution of dispersion with quantum geometry---}We now analyze the dependence of the anyon dispersion on the strength and shape of the non-uniform quantum geometry, encoded in the
K\"ahler potential. At uniform quantum geometry ($s_1 = s_2 =0 $), anyons are dispersionless. Deviation due to a weak K\"ahler potential can be accounted for perturbatively. The first order is non-vanishing and lifts the degeneracy between the Bloch states. In this limit, the bandwidth $\Delta E$ of the anyon dispersion $E_{\boldsymbol{k}}$, Eq.~\eqref{eq:energy_defn}, exhibits linear scaling with $s_n$, consistent with our numerical results in Figure~\ref{fig:bandwidth}(c)/(d) where we plot the dispersion bandwidths as a function of $s_{1/2}$ and fixed $s_{2/1}$. Moreover, the bandwidth saturates at large $s_n$. For $s_n \rightarrow \infty$, the electrons occupy only the discrete minima of the K\"ahler potential and a universal anyon dispersion emerges, setting the saturation values.

For a purely second harmonic modulation ($s_1 =0\,, s_2 = 1$), shown in Figure \ref{fig:bandwidth}(a)-(b), the dispersion is strongly suppressed compared to the first harmonic case, and exhibits an enhanced degeneracy. This behavior can be understood from the emergence of additional symmetry operators corresponding to translations of the anyons by half a unit cell along both directions. Acting on the anyon Bloch basis, these symmetries generate an additional fourfold degeneracy, rendering the spectrum $4m$-fold degenerate in the anyon magnetic BZ ($4m^2$-fold degenerate in the electronic BZ). More generally, the $n$th harmonic induces an $mn^2$-fold degeneracy in the anyon magnetic BZ ($m^2 n^2$-fold in the electronic Brillouin zone) \suppRef{as shown in}{app:additional_symmetries}. Physically, higher harmonics drive the system toward a more uniform electronic distribution, akin to the effect of a constant magnetic field \suppRef{as detailed in}{app:infinite_sn_limit}. We numerically find this suppression to be already significant at the level of the second harmonic.

At finite $s_1$, a positive (negative) value of $s_2$ leads to a reduced (enhanced) bandwidth as illustrated in Fig.~\ref{fig:bandwidth}(c)/(d). 
This can be understood by noting that a positive $s_2$ localizes electrons at the nodes of the first harmonic, driving the system to a second harmonic dominated regime and reducing the bandwidth\suppRef{, as illustrated in}{app:symmetries_rectangular}. A negative $s_2$ on the other hand reinforces the first harmonic minima, thus increasing the bandwidth. In the large $s_1$ limit however, the influence of $s_2$ becomes negligible, as the $s_1$ contributions dominate.

\emph{Discussion---} FCI anyons exhibit a non-trivial dispersion due to the non-uniform quantum geometry of the underlying Chern band. We construct anyon Bloch wavefunctions, which enable efficient computation of this dispersion. The spectrum exhibits an $m-$fold degeneracy in the anyon magnetic BZ, which we explicitly show to arise from the topological degeneracy of the FCI state. Relating the anyon and electron translation operators, we explain the $m^2-$fold degeneracy reported in earlier works to be a  splicing of states in the anyon magnetic BZ into the electronic BZ. Moreover, the anyon dispersion of the exact Coulomb ground state shows strong agreement with our trial wavefunctions. The dispersion differs only by an overall scaling up to $s_1 \sim 4$ \suppRef{as shown in}{sup:ED}, while for $s_1 >4$, the groundstate is no longer an FCI \cite{shi2025effectsberrycurvatureideal}.

We find that the anyon dispersion is governed by the strength and the structure of quantum geometry non-uniformity. The bandwidth grows linearly with weak modulation, saturating at large modulation. Remarkably, higher harmonics suppress the dispersion; our analytical framework links this to emergent symmetries where the $n$th harmonic induces an $mn^2$-fold degeneracy, driving the system toward a uniform electron distribution. Our numerical results are presented for a simple cosine form of the quantum geometry for clarity; incorporating the single-particle quantum geometry of realistic moir\'e systems such as twisted bilayer graphene~\cite{tarnopolsky19} or $\text{MoTe}_2$~\cite{crepel2024chiral} is a straightforward extension within our framework.

Our work demonstrates a high degree of analytical control over anyon excitations of FCI states. While we have focused here on the energy dispersion of one anyon, the analytical formalism developed here is quite general and paves the way for several future investigations. In the pursuit of exotic itinerant anyonic phases~\cite{Pichler_2025, nosov2025anyonsuperconductivityplateautransitions, shi2025anyondelocalizationtransitionsdisordered, zhang2025holonmetalchargedensitywavechiral,shi2025nonabeliantopologicalsuperconductivitymelting, lotric2026phasesitinerantanyonslaughlins}, the topology of anyon Bloch bands may play an important role. Beyond single anyons, our formalism could be adapted to address the effect of anyon dispersion on interactions between multiple anyons~\cite{Gattu2025, Xu2025, wang2026anyonmoleculesfractionalquantum, li2026boundstatesanyonsgeometric}.  Finally, an extension of the present analytical work to higher Chern bands~\cite{wang2022, dong2022, Eslam_highC_idealband, wang2023origin, guercimao2024, datta2024}  and higher vortexable bands~\cite{liuwang2025, Fujimoto2025}, where non-Abelian anyons may arise, presents an exciting direction for exploring emergent phases of itinerant anyons in topologically ordered lattice systems.

\textit{Acknowledgements}.---K.I. thanks Dalin Bori\c ci for insightful discussions on FQHE physics and MC numerics, and Khalak Mahadeviya for helpful inputs on the numerics. N.R. and A.F. are grateful to Miguel Gon\c{c}alves, Felipe Mendez-Valderrama, Jonah Herzog-Arbeitman, and Nicolas Morales-Duran for fruitful collaborations on related topics, and for helpful comments.  We acknowledge insightful discussion with Ajit Balram. The Flatiron Institute is a division of the Simons Foundation. This work is supported by the Deutsche Forschungsgemeinschaft (DFG, German Research Foundation) through Project-ID 258499086 -- SFB 1170 and through the W\"urzburg-Dresden Cluster of Excellence on Complexity and Topology in Quantum Matter -- ctd.qmat Project-ID 390858490 -- EXC 2147. 
Cette recherche a été financée par le Conseil de recherches en sciences naturelles et en génie du Canada (CRSNG) à travers le programme de subventions à la découverte RGPIN-2026-07726. 

\bibliography{references}

\clearpage
\onecolumngrid
\appendix

\setcounter{equation}{0}
\setcounter{figure}{0}
\setcounter{table}{0}
\setcounter{section}{0}
\setcounter{secnumdepth}{3} 
\counterwithout{equation}{section}

\makeatletter

\def\@hangfrom@section#1#2#3{#1#2:\hspace{0.5em}#3}

\renewcommand{\thesection}{\Alph{section}}
\makeatother

\renewcommand{\thesubsection}{\arabic{subsection}}
\renewcommand{\theequation}{S\arabic{equation}}
\renewcommand{\thefigure}{S\arabic{figure}}

\newcommand{\red}[1]{\textcolor{red}{#1}}
\newcommand{\blue}[1]{\textcolor{blue}{#1}}
\newcommand{\delb}[1]{\textcolor{blue}{\sout{#1}}}
\newcommand{\del}[1]{\textcolor{red}{\sout{#1}}}

\renewcommand{\theequation}{S\arabic{equation}}
\renewcommand{\thefigure}{S\arabic{figure}}
\renewcommand{\bibnumfmt}[1]{[S#1]}

\onecolumngrid

\section*{SUPPLEMENTAL MATERIAL}

\section{Ideal band single quasihole wavefunction on a torus}\label{app:wavefunction}

We consider the Laughlin-like FCI of an ideal band  at filling fraction $\nu = 1/m$ on a torus spanned by the complex vectors $L_j$, $j \in \{1, 2\}$. The many-body wavefunction is built from a single-particle ideal band, hosting a single quasihole at location $\eta$ and $N_e$ electrons at positions $z_i = x_i +iy_i$ or $\boldsymbol{r}_i = (x_i, y_i)$ and is given by~\cite{Greiter2016, ledwith2020a, Ledwith2023}:
\begin{equation}
    \psi^\sigma[\{\boldsymbol{r_i}\}, \boldsymbol{\eta}] = \Psi_{L}^\sigma[\{\boldsymbol{r_i}\}, \boldsymbol{\eta}] \prod_{i = 1}^{N_e} \chi(\bld{r_i})
\end{equation}
with
\begin{equation}
    \Psi_{L}^\sigma[\{\boldsymbol{r_i}\}, \boldsymbol{\eta}] = A_\sigma e^{\frac{1}{4ml^2}\mathcal{B}(\bld{\eta})} e^{iK_\sigma\rnd{Z + \frac{\eta}{m}}} \prod_{\gamma = 1}^m \vartheta_1\sq{\frac{Z + \frac{\eta}{m} - Z^\gamma_\sigma}{L_1} \Bigg| \tau}   \prod_{i = 1}^{N_e} e^{\frac{1}{4l^2}\mathcal{B}(\bld{r_i})}~\vartheta_1\sq{\frac{z_i - \eta}{L_1}\Bigg| \tau}   \prod_{i<k}^{N_e}\vartheta_1\sq{\frac{z_i-z_k}{L_1}\Bigg| \tau}^m\,, 
\label{eq:single_quasihole_wavefunction}
\end{equation}
and
\begin{equation}
\mathcal{B}(\bld{r}) = z^2 - |z|^2 - \mathcal{K}(\bld{r})\,.    
\end{equation}
Note that we use the word "electrons" loosely, as a substitute for the (spinless) microscopic constituents, be them bosons (for $m$ even) or fermions (for $m$ odd). $\Psi_{L}^\sigma[\{\boldsymbol{r_i}\}, \boldsymbol{\eta}]$ is the Laughlin quasihole FQH wavefunction on a torus under a non-uniform magnetic field captured by the Kähler potential  $\mathcal{K}(\bld{r})$, while $\chi(\bld{r})$ is a normalized spinor. Both come from the single-particle band from which we construct the FCI ansatz.  $A_\sigma$ is a normalization factor given by
\begin{equation}
    A_\sigma = e^{\frac{iK_1L_2}{m}\sigma}  e^{i\frac{\pi \tau}{m}\sigma(\sigma-2)},
\end{equation}
which ensure that the remaining global  normalization of the wavefunction is independent of $\sigma$, while the two terms after $A_\sigma$ in Eq. \eqref{eq:single_quasihole_wavefunction} ensures that the normalization is independent of $\eta$. Throughout the text, bolded coordinates indicate $2$D vectors and unbolded ones are in $2$D complex notation. $Z = \sum_{i=1}^{N_e}z_i$ is the electron centre of mass, up to a  factor $1/N_e$. 

The K\"ahler potential creates a periodic potential that divides the system into unit cells with $N_{j}$ sites in the $j \in \{1,2\}$ direction. The number of sites is taken equal to the number of flux quanta in the system, $N_\phi = N_1 N_2$, which in the single quasihole state is $N_\phi = mN_e +1$. This can be seen by counting the number of zeros for any electron coordinate $z_i$ in Eq.~\eqref{eq:single_quasihole_wavefunction}, which, by the Riemann-Roch theorem, equals the number of linearly independent states in the one-body Hilbert space. Hence, by construction, each unit cell encloses a single flux quantum:
\begin{equation}
    2\pi l^2 = \text{Im}(l_2\bar{l}_1),
    \label{eq:magnetic_length_definition}
\end{equation}
where $l_{j} = L_{j}/N_{j}$, and $l$ is the magnetic length. We have chosen an orientation such that $\text{Im}(l_2\bar{l}_1) > 0$, and the symmetric gauge $\bld{A} = B\bld{r}\times\bld{\hat{z}}/2$ is adopted. The K\"ahler potential $\mathcal{K}$ is taken such that its integral over a single unit cell vanishes, preserving the condition of a single flux quantum per unit cell. Continuing with the definitions,  $\tau = L_2/L_1$, while $K_\sigma$ and $Z_\sigma^\gamma$ are complex numbers
\begin{equation}
\begin{split}
    Z_\sigma^{\gamma} &= \rnd{\frac{Z_1}{m} + \frac{\gamma-1}{m}}L_1 - \frac{\sigma-1}{m}L_2\,, \\
    K_\sigma &= K_1 + \frac{2\pi}{L_1}(\sigma-1)\,,
\end{split}
    \label{eq:constraint_2}
\end{equation}
where $\sigma \in \{1,\dots,m\}$ indexes the topological sector. $K_1$, $Z_1$ are fixed implicitly by electron translations. Indeed, translating an electron over the entire system, we impose that the wavefunction satisfies the boundary conditions 
\begin{equation}
    \psi^\sigma[\{\bld{r_1},\dots,\bld{r_j}+\bld{L_{1/2}},\dots, \bld{r_{N_e}}\}, ~\boldsymbol{\eta}] = e^{i\phi_{1/2}}\psi^\sigma[\{\boldsymbol{r_i}\}, \boldsymbol{\eta}]\,,
    \label{eq:twist_def}
\end{equation}
picking up the phases $\phi_{1/2}$. Using the quasiperiodic properties of the Jacobi elliptic theta function:
\begin{equation}
\begin{split}
    \vartheta_1 (z+1|\tau) &= -\vartheta_1 (z|\tau)\,, \\
    \vartheta_1 (z+\tau|\tau)&= -e^{-i\pi\tau -2\pi i z}\vartheta_1 (z|\tau)\,.
\end{split}  
\label{eq:theta_relations}
\end{equation}
on the left hand side of Eq. \eqref{eq:twist_def}, we arrive at the following constraints on  $K_1$, $Z_1$ 
\begin{equation}
\begin{split}
    e^{iK_1 L_1} &= (-1)^{N_\phi} e^{i\phi_1}\,,\\
    e^{iK_1 L_2} e^{2\pi i Z_1} &= (-1)^{N_\phi+m-1} e^{i\phi_2}\,.
\end{split}    
\label{eq:topology_constraints}
\end{equation}
 By construction, $K_1 \in [0, 2\pi) $ for even $N_\phi$, and $K_1 \in [-\pi, \pi)$ when $N_\phi$ is odd. 
 As we show in Appendix \ref{app:topological_degeneracy}, these constraints encode the topological degeneracy of the FQH system. 

Finally, we point out an inherent incommensurability between the number of unit cells in either direction, and the filling fraction of the system.  Since $N_\phi = m N_e +1 = N_1N_2$, by definition,
\begin{equation}
    N_\phi =  N_1 N_2 =  1 ~\text{mod}~ m\,,
\end{equation}
and hence,
\begin{equation}
    \text{gcd}(N_1, m) = \text{gcd}(N_2, m) = 1\,.
    \label{eq:GCD}
\end{equation}
While these relations are relatively less important in the thermodynamic limit, they do play an important role in finite size systems to which we are numerically restricted.

\section{Electron translation operators and anyon magnetic translation operators}\label{app:translations}

Translation operators move a particle from one position to another and, in a homogeneous system, commute with the Hamiltonian. However, for charged particles even in a uniform magnetic field, ordinary  translations fail to commute with the Hamiltonian, since the particles acquire an Aharonov–Bohm phase when they move. This issue is resolved by defining magnetic translation operators, which combine a spatial translation with a compensating phase factor.

As explained in the main text, while electrons in a zero-field FCI experience no net magnetic field, anyons are \textit{inherently} magnetic entities. Their motion accumulates a many-body Berry phase in real space, effectively mimicking the role of a magnetic flux. Hence, we define standard translation operators for electrons, and magnetic translation operators for the quasihole.

For electrons, the translation operator takes the form
\begin{equation}
T_{\bld{r_j}}(\bld{a}) = e^{a\partial_{z_j} + \bar{a}\bar{\partial}_{z_j}},
\end{equation}
where the exponential differential operator translates the $j-$th electron coordinate, acting on the wavefunction as:
\begin{equation}
    T_{\bld{r_j}}(\bld{a})\psi^\sigma[\{\boldsymbol{r_i}\}, \boldsymbol{\eta}] = \psi^\sigma[\{\boldsymbol{r_1},\dots,\boldsymbol{r_j}+\boldsymbol{a},\dots,\boldsymbol{r_{N_e}} \},\boldsymbol{\eta}].
\end{equation}
Under this translation, the magnitude of the wavefunction picks up an Aharanov-Bohm phase, which is compensated by an equal and opposite phase coming from the spinor when the translations are multiples of unit cell lengths \cite{Guerci2025}. Similarly, we define quasihole magnetic translation operators from the way they act on the quasihole wavefunction:
\begin{equation}\label{eq:magnetic-translation}
   T_{\bld{\eta}}(\bld{a})\psi^\sigma[\{\boldsymbol{r_i}\}, \boldsymbol{\eta}] = e^{-\frac{i}{2ml^2}\text{Im}(\eta \bar{a})}\psi^\sigma[\{\boldsymbol{r_i}\}, \boldsymbol{\eta+a}] \,.
\end{equation} 
Under this operation, the quasihole is translated, and the wavefunction is multiplied by a compensating phase factor where the symmetric gauge is assumed.  The factor $1/m$ in the phase reflects that a quasihole carries a charge of $1/m$ times that of the electron. 

Importantly, and in contrast to the electron translation operators, the quasihole magnetic translation operators are defined only within the subspace spanned by the quasihole wavefunctions given in Eq.~\eqref{eq:single_quasihole_wavefunction}. This has two main consequences: (i) they are not defined outside the quasihole subspace, and (ii) they act exclusively on quasihole wavefunctions, not on other functions that may depend on $\eta$, such as the normalization factor in Eq.~\eqref{eq:quasihole_momentum_eigenfunction} that we will discuss in Appendix~\ref{app:anyonbloch}.
Within the subspace of quasihole functions, the quasihole magnetic translation operators (MTO) satisfy the commutation relation:
\begin{equation}
   T_{\bld{\eta}}(\bld{a})T_{\bld{\eta}}(\bld{b}) = e^{\frac{i}{2ml^2}\text{Im}(b \bar{a}-a\bar{b})}~T_{\bld{\eta}}(\bld{b})T_{\bld{\eta}}(\bld{a}). 
   \label{eq:anyon_MTO_commutator}
\end{equation}
Note that the projective phase gained by commuting the operators is independent of $\bld{\eta}$, depending only on the translations.

Let us now define quasihole translation operators over a \textit{unit cell}: 
\begin{equation}
 T_{j}\equiv T_{\bld{\eta}}(\bld{l_j}) .     
\end{equation}
It is easy to show using Eqs.~\eqref{eq:magnetic_length_definition} and ~\eqref{eq:anyon_MTO_commutator} that acting on the relevant subspace of wavefunctions, these satisfy the following projective  commutation relation:
\begin{equation}
    T_1 T_2 = e^{\frac{2\pi i}{m}}T_2T_1.
\end{equation}
Similarly, we define electron center of mass translation operators:
\begin{equation}
    \tilde{T}_{j} \equiv \prod_{i=1}^{N_e} T_{\bld{r_i}}(\bld{l_j})\,,
\end{equation}
which commute with each other. As we will show in Appendix~\ref{app:electron_quasihole_translation_relation}, these operators are related to each other by (up to phases):
\begin{equation}
    \tilde{T}_{j} = \rnd{T_{j}}^{q_{j}N_{j}-1}\,,
    \label{eq:electron_quasihole_momenta_relation}
\end{equation}
where
\begin{equation}
(q_{1}, q_{2}) = (N_{2}, N_1)~\text{mod}~m.
\label{eq:q_relations}
\end{equation}

\section{Quasihole translations and topological degeneracy on the torus}\label{app:topological_degeneracy}

The topological degeneracy of the FCI state on the torus can be deduced by considering translations of the quasihole coordinate over the two arms of the torus
\begin{equation}
\mathcal{T}_j \equiv T_j^{N_j}\label{app:eq:fulltranslationquasihole} \,.
\end{equation}
We follow Ref.~\cite{Greiter2016}, writing out some of their derivations in more detail. First, we consider a full translation of the quasihole along the $1$ direction:
\begin{equation}
\begin{split}
    \mathcal{T}_1\psi^\sigma[\{\boldsymbol{r_i}\}, \boldsymbol{\eta}] &=  A_\sigma e^{-\frac{i}{2ml^2}\text{Im}\rnd{\eta\bar{L}_1}}e^{\frac{1}{4ml^2}\mathcal{B}(\bld{\eta+L_1})} e^{iK_\sigma \rnd{Z + \frac{\eta}{m} + \frac{L_1}{m}}}\\
        &\prod_{\gamma = 1}^m \vartheta_1\sq{\frac{Z + \frac{\eta}{m} - Z^\gamma_\sigma + \frac{L_1}{m}}{L_1} \Bigg| \tau} \prod_{i = 1}^{N_e}\vartheta_1\sq{\frac{z_i - \eta -L_1}{L_1}\Bigg| \tau}\times \cdots\,,
\end{split}
\end{equation}
where we have omitted the terms that remain unaffected under quasihole translations. To absorb the $\eta$ shift in the centre of mass term, we realize that $Z^\gamma_\sigma - L_1/m = Z^{\gamma -1}_\sigma$ for $\gamma < m$ and $Z^1_\sigma - L_1/m = Z^{m}_\sigma-L_1$, allowing to relabel the product and use Eq.~\eqref{eq:theta_relations} to arrive at:
\begin{equation}
\begin{split}
\mathcal{T}_1\psi^\sigma[\{\boldsymbol{r_i}\}, \boldsymbol{\eta}] = e^{\frac{iK_\sigma L_1}{m}}(-1)^{N_e+1}\psi^\sigma[\{\boldsymbol{r_i}\}, \boldsymbol{\eta}]. 
\end{split}
\label{eq:single_translation_1}
\end{equation}
 Thus, the single quasihole coherent state-like wavefunction on the torus is an eigenfunction of the $\mathcal{T}_1$ operator. Under $m$ translations along the $1$ direction, iterating the same $m$ times and using Eq.~\eqref{eq:topology_constraints}, we have:
\begin{equation}
\mathcal{T}_1^m  \psi^\sigma[\{\boldsymbol{r_i}\}, \boldsymbol{\eta}]  = (-1)^{m-1}e^{i\phi_1}\psi^\sigma[\{\boldsymbol{r_i}\}, \boldsymbol{\eta}].
\label{eq:m_translation_1}
\end{equation}
Now, we consider a full translation along the $2$ direction:
\begin{equation}
\begin{split}
\mathcal{T}_2\psi^\sigma[\{\boldsymbol{r_i}\}, \boldsymbol{\eta}] &= A_\sigma e^{-\frac{i}{2ml^2}\text{Im}\rnd{\eta\bar{L}_2}}e^{\frac{1}{4ml^2}\mathcal{B}(\bld{\eta+L_2})} e^{iK_\sigma \rnd{Z + \frac{\eta+L_2}{m} }}\\
        &\prod_{\gamma = 1}^m \vartheta_1\sq{\frac{Z + \frac{\eta}{m} - Z^\gamma_\sigma + \frac{L_2}{m} }{L_1} \Bigg| \tau} \prod_{i = 1}^{N_e}\vartheta_1\sq{\frac{z_i - \eta+L_2}{L_1} \Bigg| \tau}\times \cdots\,.
\end{split}
\end{equation}
The shift in $\eta$ is absorbed by noting that  $Z_\sigma ^\gamma - L_2/m = Z^\gamma_{\sigma+1}$. As mentioned earlier, Eq.~\eqref{eq:constraint_2}
demands a simultaneous change: $K_\sigma \rightarrow K_{\sigma+1}  $. Further using Eq.~\eqref{eq:theta_relations} allows us to write, for $\sigma< m$:
\begin{equation}
    \mathcal{T}_2\psi^\sigma[\{\boldsymbol{r_i}\}, \boldsymbol{\eta}] = \psi^{\sigma+1}[\{\boldsymbol{r_i}\}, \boldsymbol{\eta}]\,,
    \label{eq:T2_translation_1}
\end{equation}
while for $\sigma  = m$,
\begin{equation}
    \mathcal{T}_2\psi^m[\{\boldsymbol{r_i}\}, \boldsymbol{\eta}] = (-1)^{m-1}e^{i\phi_2}\psi^{1}[\{\boldsymbol{r_i}\}, \boldsymbol{\eta}]\,.
\label{eq:T2_translation_2}
\end{equation}
For the sake of clarity, acting $\mathcal{T}_1$ on the $\mathcal{T}_2$ translated wavefunction yields:
\begin{equation}
\mathcal{T}_1\mathcal{T}_2\psi^\sigma[\{\boldsymbol{r_i}\}, \boldsymbol{\eta}] = e^{\frac{iK_{\sigma+1} L_1}{m}}(-1)^{N_e+1}\mathcal{T}_2\psi^\sigma[\{\boldsymbol{r_i}\}, \boldsymbol{\eta}],
\end{equation}
which from Eq.~\eqref{eq:constraint_2} implies an extra $e^{2\pi i /m}$ phase compared to Eq.~\eqref{eq:single_translation_1}, confirming the rotation of the topological sector by $\mathcal{T}_2$. Finally, Eqs.~\eqref{eq:T2_translation_1} and~\eqref{eq:T2_translation_2} together give us 
\begin{equation}
    \mathcal{T}_2^m\psi^\sigma[\{\boldsymbol{r_i}\}, \boldsymbol{\eta}] = (-1)^{m-1}e^{i\phi_2}\psi^{\sigma}[\{\boldsymbol{r_i}\}, \boldsymbol{\eta}]\,,
    \label{eq:m_translation_2}
\end{equation}
telling us that
$m$ translations in the $2$ direction brings the system back to the same state.

\section{Anyon Bloch wavefunctions}\label{app:anyonbloch}

In this Appendix, we define the momentum eigenfunctions of the quasihole magnetic translation operators. Under a non-uniform K\"ahler potential, these eigenfunctions could have different interaction energies leading to a non-trivial dispersion of anyons, which is forbidden by symmetry in a uniform magnetic field. Since $T_1$ and $T_2^m$ commute, a natural momentum eigenfunction takes the form:
\begin{equation}
    \Psi_{k_1k_2}[\{\bld{r_i}\}] = \frac{1}{\mathcal{N}(\bld{\eta})}\sum_{n_1=0}^{N_1-1}\sum_{n_2=0}^{N_2-1} e^{-2\pi i (k_1n_1 + m k_2n_2)} ~T_1^{n_1}T_2^{mn_2}~\psi^\sigma[\{\boldsymbol{r_i}\}, \boldsymbol{\eta}] 
\label{eq:quasihole_momentum_eigenfunction}
\end{equation}
where $k_{1}, k_{2}$ are the momenta associated with translation operators $T_1$, $T_2^m$ respectively. $\mathcal{N}(\bld{\eta})$ is an $\bld{\eta}$ dependent normalization factor that gets rid of all $\eta$ dependence on the LHS. Indeed, as mentioned in Appendix~\ref{app:wavefunction}, the number of independent states in the single quasihole Hilbert space is $N_\phi = N_1 N_2$. The Bloch states are independent of $\eta$ since for well defined momenta, the above states are uniquely defined (up to phases) and there are exactly $N_\phi$ of them.  On the other hand, while we have dropped the $\sigma$ index on the LHS, the dependence on $\sigma$ is implicit in $k_1$ as shown in the next subsection.  
We defer the proof of the orthogonality of these Bloch wavefunctions to Appendix~\ref{app:electron_quasihole_translation_relation}, where we establish a precise connection between electron and quasihole magnetic translations.

\subsection{Action of $T_1$ and $T_2^m$}\label{app:commuting_translations}

Before applying the magnetic translation operators, it is useful to clarify the structure of Eq.~\eqref{eq:quasihole_momentum_eigenfunction}.
The action of powers of $T_1$ and $T_2^m$ on $\psi^\sigma[{\boldsymbol{r_i}}, \boldsymbol{\eta}]$ generates a set of $N_\phi$ quasihole wavefunctions of the form $\psi^\sigma[{\boldsymbol{r_i}}, \boldsymbol{\eta} + n_1 \boldsymbol{l}_1 + n_2 m \boldsymbol{l}_2]$. Equation~\eqref{eq:quasihole_momentum_eigenfunction} can thus be viewed as an expansion over this basis.
As already noted in Appendix~\ref{app:translations}, the quasihole magnetic translation operators act solely on the quasihole wavefunctions. When applied to the anyon Bloch function in Eq.~\eqref{eq:quasihole_momentum_eigenfunction}, they translate each quasihole wavefunction individually according to Eq.~\eqref{eq:magnetic-translation}. The expansion coefficients, however, and in particular the normalization factor $\mathcal{N}(\boldsymbol{\eta})$, remain unchanged.

Acting with $T_1$ on the above wavefunction, the crucial step is to fold the $n_1 = N_1$ contribution back to $n_1 = 0$. This is possible when the constraint
\begin{equation}
\begin{split}
    e^{-2\pi i k_1 N_1}e^{\frac{iK_1 L_1}{m}}e^{\frac{2\pi i}{m}(\sigma-1)}(-1)^{N_e+1
    } &= 1
\\
\end{split}
\label{eq:k1_constraint}
\end{equation}
is satisfied, where we use the action of $\mathcal{T}_1$ on the single quasihole wavefunction given in Eq.~\eqref{eq:single_translation_1}. Note that $K_1$ is related to the flux $\phi_1$ via Eq.~\eqref{eq:topology_constraints}. Interestingly, the allowed values of the $k_1$ momenta depend on the topological sector $\sigma$. Similarly, under the action of $T_2^m$, the constraint that needs to be obeyed is:
\begin{equation}
        e^{-2\pi i k_2 m N_2}(-1)^{m-1}e^{i\phi_2} = 1. \label{eq:k2_constraint}
\end{equation}
where we use the action of $\mathcal{T}_2^{m}$ on the localized single quasihole wavefunction, given in Eq.~\eqref{eq:m_translation_2}. As long as these constraints are satisfied, under the action of unit cell translations, the anyon Bloch wavefunctions exhibit the following quasiperiodic behavior
\begin{equation}
\begin{split}
     T_1  \Psi_{k_1k_2}[\{\bld{r_i}\}] &= e^{2\pi i k_1 }  \Psi_{k_1k_2}[\{\bld{r_i}\}],\\
         T_2^m  \Psi_{k_1k_2}[\{\bld{r_i}\}] &= e^{2\pi i mk_2 }  \Psi_{k_1k_2}[\{\bld{r_i}\}].
\end{split}
\label{eq:translation_eigenvalues}
\end{equation}
Using Eqs. \eqref{eq:k1_constraint}-\eqref{eq:k2_constraint}, we can write the set of allowed momenta to be:
\begin{equation}\label{eq:anyon-momenta}
\begin{split}
k_1 & = \frac{n_1}{N_1} + \frac{1}{2\pi N_1} 
\begin{cases} 
 \frac{\phi_1}{m} + \frac{2\pi (\sigma-1)}{m} + \pi~\text{mod}(m-1,2)  & N_\phi \in \text{even} \\
 \frac{\phi_1}{m} + \frac{2\pi (\sigma-1)  }{m} + \frac{\pi}{m}+\pi~\text{mod}(m-1,2)  & N_\phi \in \text{odd}
\end{cases} , \quad n_1 \in \{0, \dots, N_1-1\} \\
k_2 & = \frac{n_2}{m N_2} + \frac{1}{2\pi m N_2} \rnd{\phi_2 - \pi~\text{mod}(m-1,2)}\,, \quad n_2 \in \{0, \dots, N_2-1\},
\end{split}
\end{equation}
such that $k_1 \in [0, 1)$ and $k_2 \in [0, 1/m)$, where we have used Eq. \eqref{eq:topology_constraints}.

\subsection{Action of $T_2$}\label{app:T2}
We now ask what the action of the operator $T_2$ on the quasihole translation eigenfunctions is:
\begin{equation}
\begin{split}
T_2\Psi_{k_1k_2}[\{\bld{r_i}\}] = \sum_{n_1=0}^{N_1-1}\sum_{n_2=0}^{N_2-1} e^{-2\pi i (n_1(k_1+1/m) + m k_2n_2)} ~T_1^{n_1}T_2^{mn_2+1} ~\psi^\sigma[\{\boldsymbol{r_i}\}, \boldsymbol{\eta}]    
\end{split}
\end{equation}
where we have used the commutator of $T_1, T_2$. Acting $T_1$ on the above wavefunction, we find:
\begin{equation}
    T_1T_2\Psi_{k_1k_2} =  e^{2\pi i \rnd{k_1 +1/m}}T_2\Psi_{k_1k_2}
\end{equation}
revealing that 
\begin{equation}
    T_2 \Psi_{k_1,k_2} = \Psi_{k_1+1/m,k_2}\,.
    \label{eq:k1_shift_T2}
\end{equation}
The same constraints as Equation~\eqref{eq:k1_constraint} need to be satisfied to start with, for this relation to hold. However, it should be noted that in finite size systems to which we are (numerically) limited, $k_1+1/m$ does not satisfy Equation~\eqref{eq:k1_constraint} once the topological sector and boundary conditions are fixed. Still, this property underlies the $m-$fold degeneracy in the anyon dispersion, revealed by comparing the results from different boundary conditions.

\subsection{Topological origin of the $m-$fold degeneracy}\label{app:T2_topological_degeneracy}

As mentioned above, the momenta $k_1$ and $k_1 +1/m$ cannot be both consistent with Eq.~\eqref{eq:k1_constraint} in the topological sector $\sigma$. Indeed, if $k_1$ is consistent with~\eqref{eq:k1_constraint} for a topological sector $\sigma$, $k_1+1/m$  obeys Eq.~\eqref{eq:k1_constraint} for the sector $\sigma + q_1$, where $N_j = q_j\mod m$.
This is a \textit{direct} indication of the Bloch state degeneracy originating from the topological degeneracy, as hinted by TQFT arguments~\cite{cheng2016a, bultinck2018a, Shi2025}.

In the localized quasihole basis, as shown by Ref.~\cite{Greiter2016}, $\mathcal{T}_2$ rotates the system  between the $m-$fold degenerate states.  Using $N_2 = q_2 + m r'$, with $r' \in \mathbb{Z}$, we can write $T_2^{q_{2}} = \sq{e^{-2\pi i m k_2}}^{r'} \mathcal{T}_2$. Moreover, the flux-particle relation $N_\phi = m N_e+1 = N_1 N_2$ implies that $q_1 q_2 = mr'' + 1$, $r''\in \mathbb{Z}$, allowing us to finally write 
\begin{equation}
    T_2 = \sq{e^{-2\pi i m k_2}}^{r} \mathcal{T}_2^{q_1}
\end{equation}
with $r =r'q_1+r'' $, showing that the action of $T_2$ is exactly to rotate the state into a topologically degenerate one, up to phases.

\section{Mapping between quasihole and electron Brillouin zones}\label{app:electron_quasihole_translation_relation}

In this Appendix, we focus on the case $\phi_1, \phi_2 =0$, {\it i.e.} in the absence of flux insertion. We derive a relation between the electron and quasihole unit cell translation operators.We then use it to precisely describe how quasihole momenta are unfolded onto electron momenta, and to prove orthogonality between the Bloch wavefunctions introduced in Appendix~\ref{app:anyonbloch}. 

\subsection{Connecting the electron and quasihole magnetic translations}

Consider the single quasihole wavefunction on the torus defined in Eq. \eqref{eq:single_quasihole_wavefunction}, repeated here for convenience:
\begin{equation}
    \psi^\sigma[\{\boldsymbol{r_i}\}, \boldsymbol{\eta}] = ~e^{\frac{1}{4ml^2}\mathcal{B}(\bld{\eta})} \prod_{i=1}^{Ne}e^{\frac{1}{4l^2}\mathcal{B}(\bld{r_i}) }~\chi(\bld{r_i})
     e^{iK_\sigma \rnd{Z + \frac{\eta}{m}}} \prod_{\gamma = 1}^m \vartheta_1\sq{\frac{Z + \frac{\eta}{m} - Z^\gamma_\sigma}{L_1} \Bigg| \tau} \prod_{i = 1}^{N_e}\vartheta_1\sq{\frac{z_i - \eta}{L_1} \Bigg| \tau} J (\{z_i \}),
\end{equation}
where the relative terms between the electrons are included in the Jastrow factor $J (\{z_i \}) = A_\sigma  \prod_{i<k} \vartheta_1 \left( \frac{z_i-z_k}{L_1} | \tau \right)^m$, which remains unaffected by the translations applied in this Appendix.

\begin{figure}
    \centering
    \includegraphics[width=0.7\linewidth]{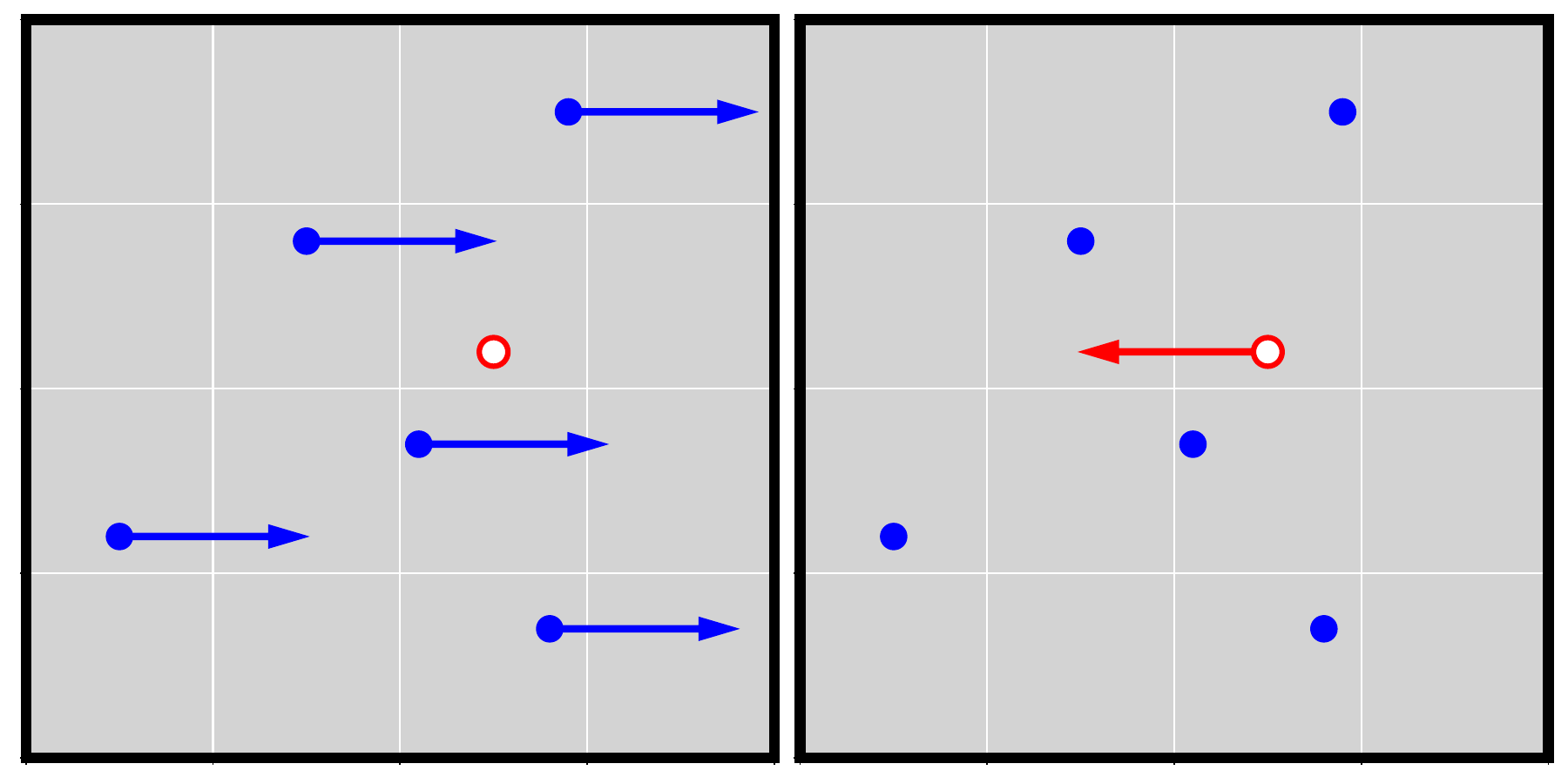}
    \caption{\textbf{Relation between electron and anyon translations.} We illustrate a system with five electrons (in blue) and one quasihole (red). On the left, all the electrons are translated by $l_1$ to the right. As shown on the right, on a torus, this is the same as translating the anyon by $-l_1$ to the left, up to translations of the anyon along the full system length $L_1$, as derived in Eq. \eqref{eq:translation_operators_relation}.}
    \label{fig:translations_illustration}
\end{figure}

We now translate the electron centre of mass by one unit cell in the $1-$direction, and the quasihole by $-(q_1N_1-1)$ unit cells in the same direction, where $q_j $ is defined in Eq.~\eqref{eq:q_relations}. We get:
\begin{equation}
\begin{split}
   &T_1^{-(q_1N_1-1)} ~\tilde{T}_1 ~\psi^\sigma[\{\boldsymbol{r_i}\}, \boldsymbol{\eta}] \\
   &=  e^{\frac{2i(q_1N_1-1)}{4ml^2}\text{Im}\sq{\eta \bar{l}_1 } }  
   e^{\frac{1}{4ml^2}\mathcal{B}[\bld{\eta}-(q_1N_1-1)\bld{l_1}]} \prod_{i=1}^{Ne}e^{\frac{1}{4l^2}\mathcal{B}(\bld{r_i+l_1}) }~\chi(\bld{r_i+l_1}) e^{iK_\sigma \rnd{Z +  \frac{\eta}{m}+\rnd{N_e- \frac{q_1N_1-1}{m}}l_1}}  \\
    & \prod_{\gamma = 1}^m \vartheta_1\sq{\frac{Z + \frac{\eta}{m} - Z^\gamma_\sigma}{L_1} + \rnd{N_e- \frac{q_1N_1-1}{m}} \frac{l_1}{L_1}\Bigg| \tau} \prod_{i = 1}^{N_e}\vartheta_1\sq{\frac{z_i - \eta}{L_1} + \frac{q_1N_1l_1}{L_1}\Bigg| \tau}\, J (\{z_i \}).
\end{split} \label{appeq_T1matching_1}
\end{equation}
By definition, the electrons do not experience an external magnetic field, so the combination $e^{\frac{1}{4l^2}{\cal B}(\bm{r})} \chi(\bm{r})$ does not pick up an Aharanov-Bohm phase and is invariant under the shift by $\bld{l}_1$. The quasihole in contrasts acquires a Berry phase $(4ml^2)^{-1} \{\mathcal{B}[\bm{\eta} -(q_1N_1-1)\bm{l_1}] - \mathcal{B}(\bm{\eta})\}$, which precisely cancels the factor introduced by the magnetic translation $(4ml^2)^{-1}(q_1N_1-1) \Im (\eta \bar{l}_1) $.
Now, turning to the second line of Eq.~\eqref{appeq_T1matching_1}, we use $N_1l_1/L_1 = 1$  to get
\begin{equation}
   \rnd{N_e- \frac{q_1N_1-1}{m}} \frac{l_1}{L_1} = \frac{1}{N_1}\frac{N_\phi -q_1N_1}{m}  = \frac{1}{N_1}\frac{N_1(N_2-q_1)}{m} = \frac{N_2-q_1}{m}
\end{equation}
By definition, $(N_2-q_1)/m \equiv c_1 \in \mathbb{Z}$, allowing us to use the quasiperiodicity relations of the theta function in Eq. \eqref{eq:theta_relations}:
\begin{equation}
    \tilde{T}_1 ~\psi^\sigma[\{\boldsymbol{r_i}\}, \boldsymbol{\eta}] = (-1)^{c_1(m+N_\phi)+q_1N_e} ~T_1^{(q_1N_1-1)} ~\psi^\sigma[\{\boldsymbol{r_i}\}, \boldsymbol{\eta}]
\end{equation}
A similar relation can be deduced for translations along the second direction, where 
\begin{equation}
    \tilde{T}_2 \, \psi^\sigma [\{\boldsymbol{r_i}\}, \boldsymbol{\eta}] = (-1)^{c_2(m+N_\phi) + q_2N_e} \, T_2^{(q_2N_2-1)} \, \psi^\sigma[\{\boldsymbol{r_i}\}, \boldsymbol{\eta}], \qquad \qquad c_2 = \frac{N_1-q_2}{m} \in \mathbb{Z}
\end{equation}
Thus, we find the following relation between the electron and quasihole translation operators: 
\begin{equation}
\tilde{T}_{j} = (-1)^{p_j} \, T_{j}^{q_jN_j-1} ,
\label{eq:translation_operators_relation}
\end{equation}
which depends on the parity $p_j = \text{mod}\sq{c_j(m+N_\phi) +q_jN_e, ~2}$. This relation is intuitive; as illustrated in Figure \ref{fig:translations_illustration}, it says that moving all electrons in one direction is the same as moving the anyon in the opposite direction (up to change in topological sector for $j=2$).

\subsection{Relation between the electron and quasihole momenta}

Using the above relation between the electron and quasihole translations, we can relate the electron momenta $\tilde{k}_j$ and quasihole momenta $k_j$ by:
\begin{equation}\label{eq:transformks}
    \tilde{k}_{j} = \sq{k_{j}(q_{j}N_{j}-1) + \frac{p_j}{2}}~\text{mod}~1\,,
\end{equation}
where tilde denotes the electron momenta, while the modulus accounts for the fact that the latter lie in the interval $[0 ,1)$ in both directions. We also remind that $q_1$ and $q_2$ are, respectively, $N_2$ and $N_1$ mod $m$, as defined in Eq.~\eqref{eq:q_relations}. The quasihole momenta $k_j$ are quantized on a grid as given by Eq.~\eqref{eq:anyon-momenta}. 
Let's unpack what the transformation Eq.~\eqref{eq:transformks} does. In the $1-$direction:
\begin{equation}\label{eq:relationk1-ktilde1}
    \tilde{k}_1 = \rnd{-k_1 + q_1\frac{\sigma-1}{m} + u_1} ~\text{mod}~ 1
\end{equation}
where
\begin{equation}
u_1 = 
\begin{cases} 
 \frac{p_1}{2} +\frac{q_1}{2} \text{mod}(m-1, 2)   & N_\phi \in \text{even} \\
  \frac{p_1}{2} +\frac{q_1}{2} \text{mod}(m-1, 2) + \frac{q_1}{2m}  & N_\phi \in \text{odd}
\end{cases} 
\end{equation}
 follows from Eq. \eqref{eq:anyon-momenta}, adding a constant shift to the momenta. This transformation is a rigid isometry on the circle  $[0, 1)$ on which $\tilde{k}_1$ is defined:  it is a reflection combined with a simple rigid translation, effectively reversing the ordering of the electronic momenta (compared to anyon momenta) up to shift. This is confirmed by the constant step size $\tilde{k}_1 [n_1+1] - \tilde{k}_1 [n_1] = -1/N_1$ (under the identification $0 \equiv 1$). Consequently, any continuous degenerate structure in terms of anyon momenta is preserved (but re-indexed) in terms of electronic momenta. On the other hand, in the $2-$direction:
\begin{equation}
\begin{split}
    \tilde{k}_2 &= \rnd{-k_2 + \frac{n_2 q_2}{m} + u_2}\text{mod}~1.
\end{split}
\label{eq:k2_anyon_electron_transformation}
\end{equation}
where $u_2 = -q_2\text{mod}(m-1,2)/(2m)$. The term $n_2 q_2/m~\text{mod}~1 $ takes one of the values
\[
0,\, \frac{1}{m},\, \dots,\, \frac{m-1}{m}.
\]
Therefore, as $n_2$ increases by one, the value of $\tilde{k}_2$ jumps between the $m$ subintervals
\[
[0,1/m),\; [1/m,2/m),\; \dots,\; [(m-1)/m,1).
\]
Neighboring $k$-points (whose energies differ only slightly in the reduced Brillouin zone) are redistributed across these $m$ domains, leading to the $m$-fold increase in degeneracy when the spectrum is plotted over the electronic Brillouin zone.

\begin{figure}[t]
    \centering
    \includegraphics[width=0.9\linewidth]{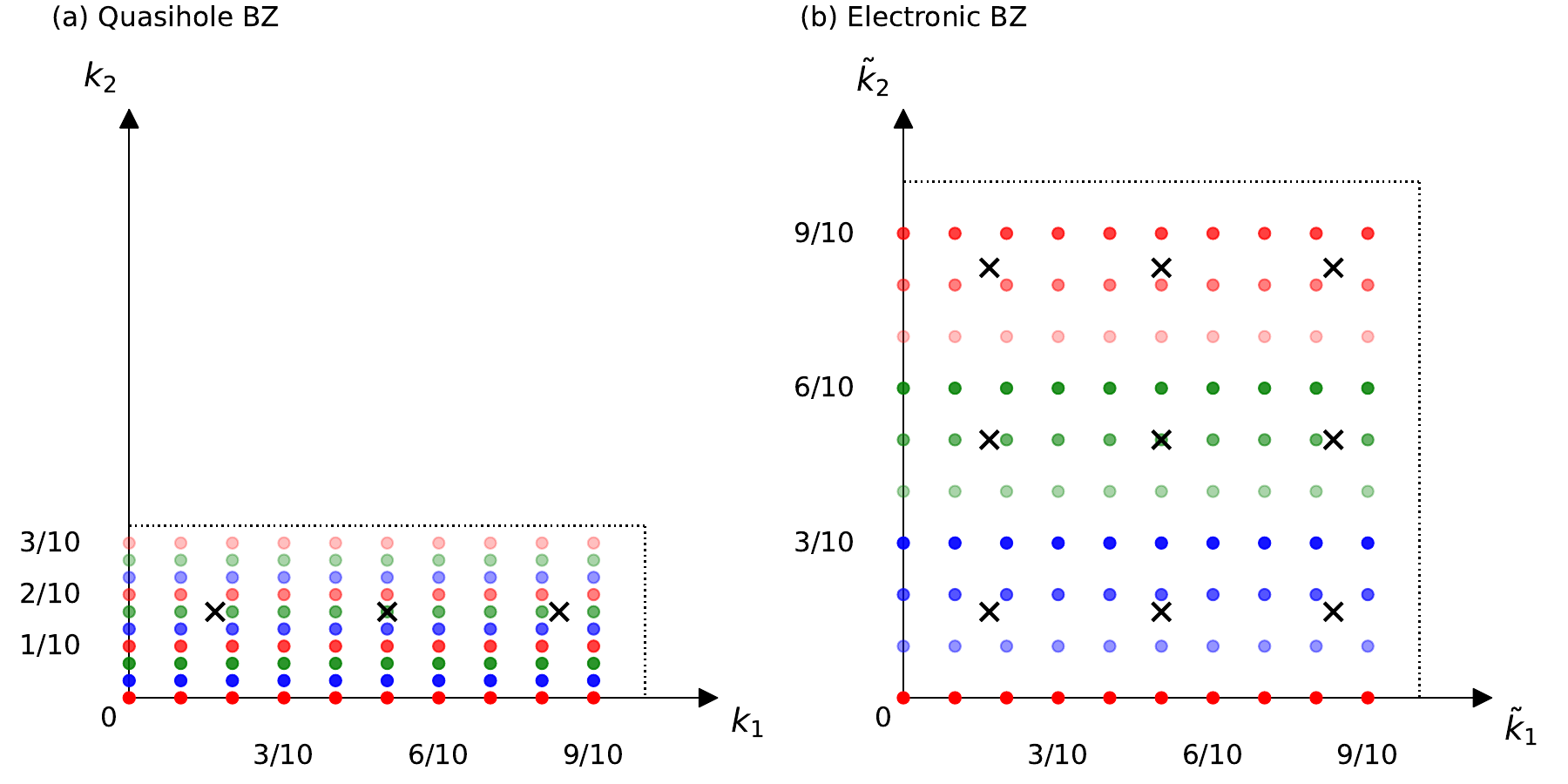}
    \caption{\textbf{Relation between anyon and electron momenta}. Relation between the quasihole momentum (left panel) and the electronic momentum (right panel). Here we consider a $N_1\times N_2=10 \times 10$ grid for $m=3$ with $\phi_1=\phi_2=0$. On the left panel, dots are colored red, blue, and green based on $n_2 ~\text{mod}~ 3$ being 0, 1, or 2, respectively. The transformation from the quasihole momentum to the electronic momentum is obtained using Eq. \eqref{eq:k2_anyon_electron_transformation}. The black crosses point the location of the anyon dispersion maxima. Note that only one of three (nine) maxima in the quasihole (electronic) Brillouin zone can be exactly obtained with such a momentum mesh. The remaining maxima can be obtained using twisted boundary conditions and a proper choice of $\phi_1$ and $\phi_2$ (see App.~\ref{app:flux_insertion}). 
}  \label{app:fig:bzconversion}
\end{figure}

As an example, for $m =3$ and $p_2 = 0$:
\begin{itemize}
    \item For $n_2 ~\text{mod}~3 = 0$: 
    \begin{equation}
        \tilde{k}_2 = \rnd{-\frac{n_2}{3N_2}}~\text{mod}~1\label{app:eq:mod_0_conv}
     \end{equation}
    Since $n_2/(3N_2) \in [0,1/3)$, these values of $\tilde{k}_2$ lie in the interval $[2/3,1)$ for $n_2 > 0$ while $n_2=0$ stays at zero.
     \item For $n_2 ~\text{mod}~3 = 1$:
         \begin{equation}
        \tilde{k}_2 = \rnd{\frac{1}{3}-\frac{n_2}{3N_2}}~\text{mod}~1.\label{app:eq:mod_1_conv}
     \end{equation}
    Since $n_2/(3N_2) \in [0,1/3)$, the above values lie in the interval $[0, 1/3)$.
     \item Finally, for $n_2 ~\text{mod}~3 = 2$:
         \begin{equation}
        \tilde{k}_2 = \rnd{\frac{2}{3}-\frac{n_2}{3N_2}}~\text{mod}~1\,,\label{app:eq:mod_2_conv}
     \end{equation}
    and by the same reasoning as above, these values lie in the interval $[1/3, 2/3)$.
\end{itemize}
This relation is illustrated in Figure~\ref{app:fig:bzconversion}.

\subsection{Orthogonality of the anyon Bloch wavefunctions}

The anyon Bloch wavefunctions in Eq.~\eqref{eq:quasihole_momentum_eigenfunction} are also eigenstates of the electron translation operators. This can be verified explicitly by acting with $\tilde{T}_1$ and $\tilde{T}_2^m$:
\begin{equation}
\tilde{T}_1 \Psi_{k_1k_2}[\{\bld{r_i}\}] = e^{2\pi i \tilde{k}_1 } \Psi_{k_1k_2}[\{\bld{r_i}\}], \qquad
\tilde{T}_2^m \Psi_{k_1k_2}[\{\bld{r_i}\}] = e^{2\pi i m \tilde{k}_2 } \Psi_{k_1k_2}[\{\bld{r_i}\}]
\end{equation}
The relations between the electron and quasihole momenta are given by
\begin{equation} 
e^{2\pi i \tilde{k}_1 } = (-1)^{N_2} e^{i K_\sigma L_1 \frac{N_2}{m}} e^{-2\pi i k_1 } \qquad \qquad e^{2 \pi i m \tilde{k}_2} = (-1)^{N_1 (N_\phi +m)} e^{-2 \pi i m k_2}
\end{equation}
 which are strictly equivalent to Eqs.~\eqref{eq:relationk1-ktilde1} and \eqref{eq:k2_anyon_electron_transformation}. Using the unitarity condition $\tilde{T}_1^\dagger \tilde{T}_1 = 1$, we obtain
 \begin{equation}
     \langle \Psi_{k_1k_2} | \Psi_{k_1' k_2'} \rangle = \langle \Psi_{k_1k_2} | \tilde{T}_1^\dagger \tilde{T}_1 | \Psi_{k_1' k_2'} \rangle = e^{2 i \pi (\tilde{k}_1'-\tilde{k}_1)}  \langle \Psi_{k_1k_2} | \Psi_{k_1' k_2'} \rangle 
 \end{equation}
It follows that the overlap must vanish unless $\tilde{k}_1 = \tilde{k}_1'$. Similarly, inserting $(\tilde{T}_2^\dagger)^m (\tilde{T}_2)^m = 1$ shows that $\langle \Psi_{k_1k_2} | \Psi_{k_1' k_2'} \rangle$ can be non-zero only if $e^{2 i \pi m (\tilde{k}_2'-\tilde{k}_2)} = 1$. Since $\text{gcd}(N_2, m) = 1$, the allowed electron momenta do not admit distinct values separated by a multiple of $1/m$, and this condition therefore implies $\tilde{k}_2' = \tilde{k}_2$. This proves that the normalized Bloch wavefunctions Eq.~\eqref{eq:quasihole_momentum_eigenfunction} form an orthonormal set
\begin{equation}
 \langle \Psi_{k_1k_2} | \Psi_{k_1' k_2'} \rangle  = \delta_{k_1,k_1'} \delta_{k_2,k_2'}    
\end{equation}

\section{Additional symmetries from higher harmonics}\label{app:additional_symmetries}

In this Appendix, we generically consider a K\"ahler potential of the form: 
\begin{equation}
 \mathcal{K}(\boldsymbol{r}_i) = \sum_{n=1}^\infty 4s_{n} \sum_{c=1,2}\frac{1}{|n\vec{b}_c|^2} \cos(n\vec{b}_c\cdot \vec{r}_i ) 
\end{equation}
where $s_n$ is the strength of the $n-$th harmonic, while $\vec{b}_c$ are the reciprocal lattice vectors of the torus. The first harmonic has the periodicity of the unit cell. If $s_n = 0$ for all $n>1$, as explained in the main text, the anyon feels the same background potential when translated by multiples of $l_{1/2}$. It follows that the operators $T_1, T_2$ commute with the Hamiltonian projected to the subspace of anyon Bloch wavefunctions. 

The higher harmonics have a greater periodicity. For clarity, we focus on a single harmonic $N>1$, i.e., $s_N \neq 0$ and $s_n = 0$ for $n\neq N$. Then, the K\"ahler potential is invariant under translations by $l_{1/2}/N$ and thus the anyon feels the same background potential under these translations. We introduce new operators:
\begin{equation}
 T^{1/N}_{j}\equiv T_{\eta}(l_{j}/N) .     
\end{equation}
that commute with the Hamiltonian. Using Eq.~\eqref{eq:anyon_MTO_commutator}, we find the following commutators of these operators with $T_1, T_2^m$:
\begin{equation}
\begin{split}
    T_1 T_2^{1/N} &= e^{\frac{2\pi i}{m N}} T_2^{1/N}T_1 \,, \\
    T_1^{1/N} T_2 &= e^{\frac{2\pi i}{mN}} T_2 T_1^{1/N}.
\end{split}    
\end{equation}
With the above commutation relations, the action of these operators on the anyon Bloch wavefunctions are:
\begin{equation}
\begin{split}
    T^{1/N}_1 \Psi_{k_1, k_2}[\{\boldsymbol{r}_i\}, \boldsymbol{\eta}] &= \Psi_{k_1, k_2+1/(mN)}[\{\boldsymbol{r}_i\}, \boldsymbol{\eta}]\\
    T^{1/N}_2 \Psi_{k_1, k_2}[\{\boldsymbol{r}_i\}, \boldsymbol{\eta}] &= \Psi_{k_1+1/(mN), k_2}[\{\boldsymbol{r}_i\}, \boldsymbol{\eta}].
\end{split}
\end{equation}
The operators $T^{1/N}_1, T^{1/N}_2$ circulate between the Bloch states labeled by momenta $\rnd{k_1 + \frac{i}{mN}, k_2+ \frac{j}{mN}}$, with $i, j \in \{0,\dots,N-1\}$. Since these operators commute with the Hamiltonian, the higher harmonics lead to an $mN^2-$fold degenerate manifold in the anyon magnetic BZ, with the degeneracy being generated by the symmetry operators $ T^{1/N}_1, T^{1/N}_2,$ and $T_2$.  From the arguments in Appendix \ref{app:electron_quasihole_translation_relation}, this  $mN^2-$fold degeneracy in the anyonic magnetic BZ will turn into a $m^2N^2-$fold degeneracy in the electronic BZ.

\section{Infinite $s_N$ limit}\label{app:infinite_sn_limit}

In this Appendix, we consider the limit of infinitely strong modulation of the K\"ahler potential
\begin{equation}
 \mathcal{K}(\boldsymbol{r}_i) = 4s_{N} \sum_{c=1,2}\frac{1}{|N\vec{b}_c|^2} \cos(N\vec{b}_c\cdot \vec{r}_i ) \,.
\end{equation}
achieved by taking $s_N \rightarrow \infty $. Within the plasma analogy of the Laughlin wavefunction, a non-zero $s_N$ breaks the translational symmetry of the uniform Laughlin plasma density to a discrete one, causing electrons to preferentially localize around the minima of the K\"ahler potential. In the limit of $s_N \rightarrow \infty $ the electrons only occupy the discrete set of minima of the K\"ahler potential. As discussed in the main text, the anyon bandwidth grows with $s_N$ but eventually saturates at a value determined by this discrete lattice limit.

\begin{figure}
    \centering
    \includegraphics[width=0.8\linewidth]{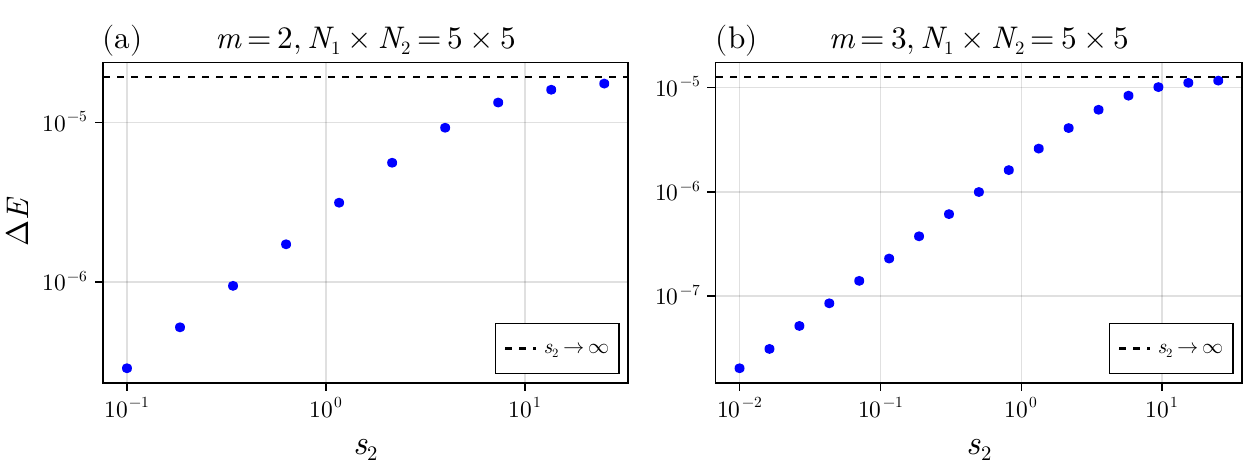}
    \caption{\textbf{Bandwidth extrapolation to infinite $\boldsymbol{s_2}$ in ED}. [(a)] $m=2$ and [(b)] $m=3$, $N_1\times N_2=5\times 5$. By linear extrapolation in $s_2^{-1}$, we obtain the saturation values of the bandwidth for this system size for $m=2$: $(1.932\pm 0.002)10^{-5}$ and $m=3$: $(1.260\pm0.005)10^{-5}$.}
    \label{fig:infinite_s2_extrapolation_ED}
\end{figure}

We present results here in the infinite $s_N$ limit using both MC and ED. This limit is directly accessible in MC by restricting electrons to occupy a finite set of points corresponding to the K\"ahler potential minima. The ED results are obtained by performing calculations for very large $s_N$ and extrapolating to the infinite $s_N$ limit, as shown in Figure \ref{fig:infinite_s2_extrapolation_ED} for $N=2$.

\begin{figure}
    \centering
    \includegraphics[width=0.8\linewidth]{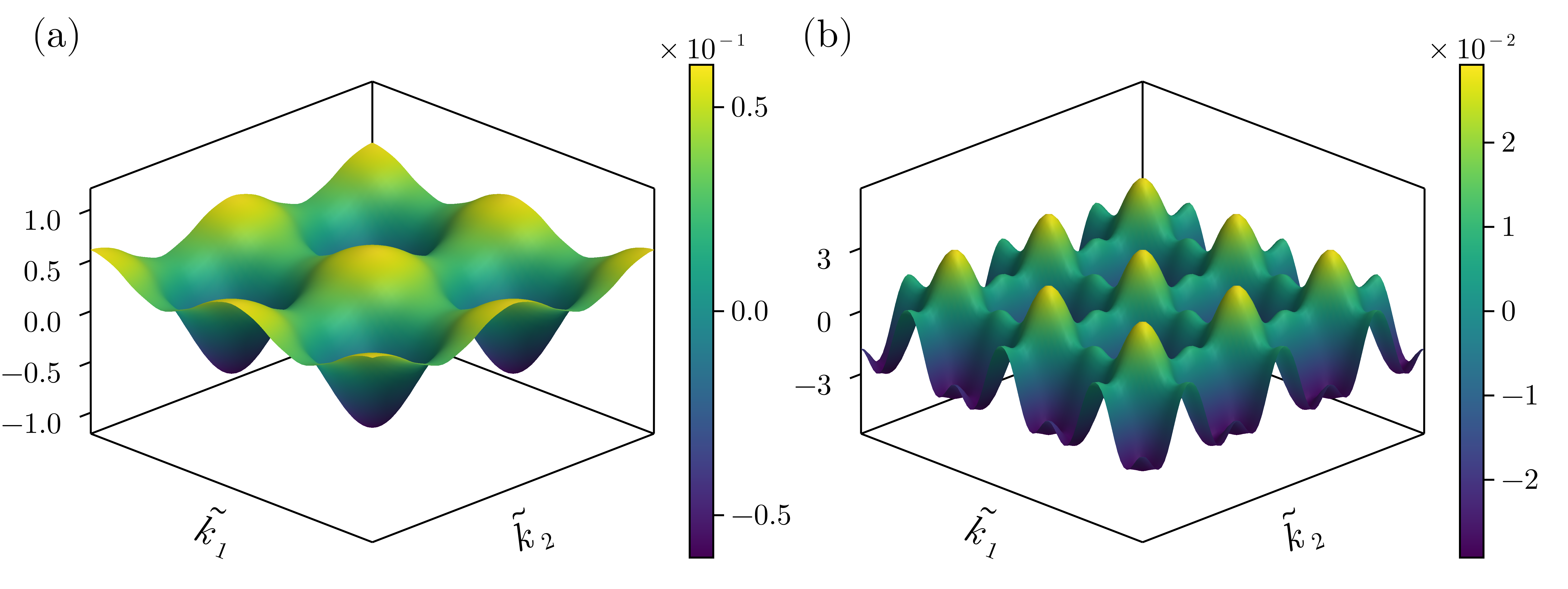}
    \caption{\textbf{Dispersion at large $\boldsymbol{s_1}$}. Dispersion for $m=2$, $5\times5$ [(a)] and $m=3$, $5\times 5$ [(b)] for $s_1=50.0$ for the anyon states in ED. Specifically on the smaller system [(b)], we observe the appearance of higher harmonics in the dispersion. We attribute this to the form factors, which depend on $\exp\rnd{-\sum_{i=1}^{N_e}\mathcal{K}(r_i)/(2l^2)}$. For large $s_1$, this introduces higher harmonics. As the anyon dispersion is generated by the band projected interaction, which depends on the form factors, this imprints the higher harmonics into the anyon dispersion.}
\label{fig:large_s1_dispersion_ED}
\end{figure}

\begin{figure}
    \centering
    \includegraphics[width=0.8\linewidth]{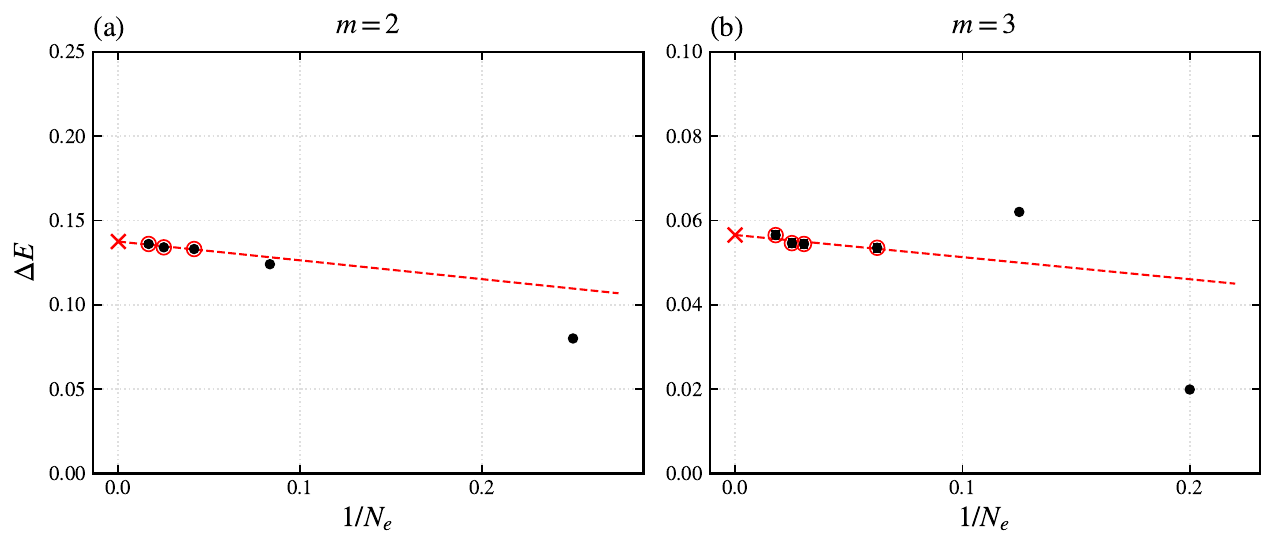}
    \caption{\textbf{Finite-size scaling of anyon bandwidth in the $\boldsymbol{s_1 \to \infty}$ limit}. Anyon bandwidths $\Delta E$ are plotted against the inverse number of electrons $1/N_e$ for (a) $m=2$ and (b) $m=3$. In both panels, black markers represent numerical data points, while red circles highlight the subset of data included in the scaling analysis. The dashed red lines denote linear fits used to extrapolate the bandwidth to the thermodynamic limit ($1/N_e \to 0$), represented by a red cross. (a) For $m=2$, we present data at system sizes $N_e \in \{4, 12, 24, 40, 60\}$, yielding an extrapolated bandwidth of $0.137 \pm 0.001$. (b) For $m=3$, we present data at $N_e \in \{5, 8, 16, 33, 40, 56\}$, yielding an extrapolated bandwidth of $0.057 \pm 0.001$ in the thermodynamic limit.}
\label{fig:finite_size_scaling_infinite_s1}
\end{figure}

In the large $s_1$ limit, in contrast to the results shown in the main text (obtained for $s_1 = 1$), we find that the anyon dispersion spectrum is richer with a higher harmonic structure as shown in Figure~\ref{fig:large_s1_dispersion_ED} for $s_1 = 50$. We attribute this to the form factors which depend on $\exp\rnd{-\sum_{i=1}^{N_e}\mathcal{K}(r_i)/(2l^2)}$. For large $s_1$, the Fourier modes of this exponential beyond the first one become significant, and are reflected in the anyon dispersion. The higher harmonic structure and the bandwidth at large $s_1$ is consistent with MC at infinite $s_1$. In Figure~\ref{fig:finite_size_scaling_infinite_s1}, we present a scaling analysis of the bandwidth at the infinite $s_1$ limit which shows that finite size effects are under control, allowing an extrapolation of the bandwidth to the thermodynamic limit.

\begin{figure}
    \centering
    \includegraphics[width=0.8\linewidth]{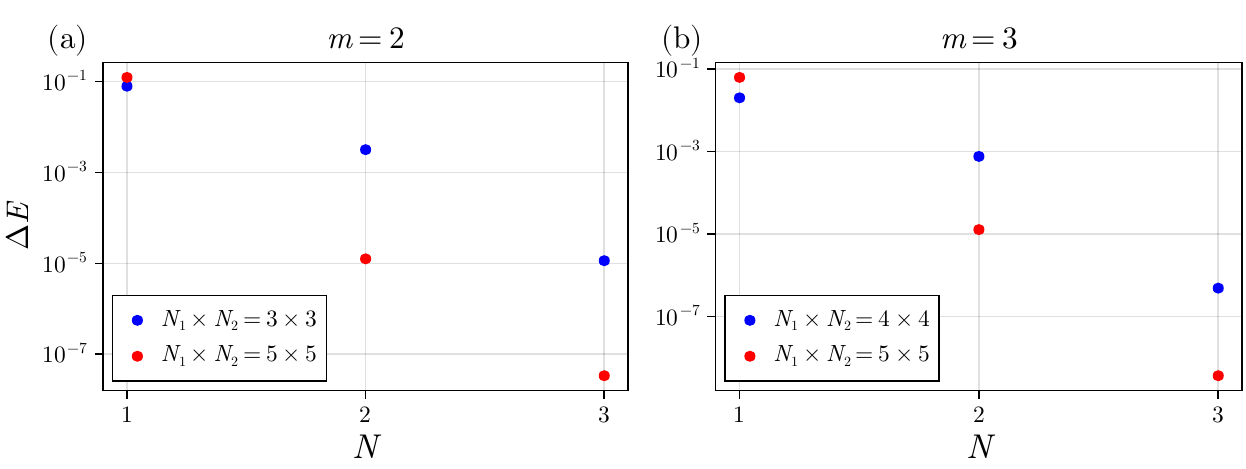}
    \caption{\textbf{Bandwidth decrease for higher harmonics}. We plot the bandwidth for $s_N\rightarrow \infty$ from ED for different system sizes and $m=2$ and $3$. We observe an exponential decay of the bandwidth vs. $N$, rendering the higher harmonics nearly irrelevant for the bandwidth compared to $N=1$. Due to strong finite-size effects, we cannot reasonably extrapolate to the saturated bandwidths in the thermodynamic limit.}
    \label{fig:large_sn_bandwidths_ED}
\end{figure}

We further quantify the dependence of the anyon dispersion on the K\"ahler potential harmonics  by analyzing the bandwidth as a function of $N$ in Figure \ref{fig:large_sn_bandwidths_ED}. The bandwidth decays exponentially with $N$, rendering the higher harmonics negligible for the bandwidth compared to $N=1$.

\section{Symmetries and interplay of K\"ahler harmonics on a rectangular lattice}\label{app:symmetries_rectangular}
First, let's just consider a single finite harmonic $s_N$ and the associated K\"ahler potential
\begin{equation}
 \mathcal{K}(\boldsymbol{r}_i) = 4s_{N} \sum_{c=1,2}\frac{1}{|N\vec{b}_c|^2} \cos(N\vec{b}_c\cdot \vec{r}_i ) \,.
\end{equation}
By inserting flux $(\phi_1, \phi_2)=(\pi/N,\pi/N)$, we take $s_N\rightarrow -s_N$. As this just corresponds to shifting the anyon dispersion, the bandwidth and the overall shape of the anyon dispersion is unaffected by this flux insertion. Therefore, we can conclude that the dispersion for a single finite $s_N$ only depends on its strength $|s_N|$, and not its sign.

Let us now consider the generic situation
\begin{equation}
 \mathcal{K}(\boldsymbol{r}_i) = 4\sum_{n=N}^{\infty} s_{n} \sum_{c=1,2}\frac{1}{|n\vec{b}_c|^2} \cos(n\vec{b}_c\cdot \vec{r}_i ) \,,
\end{equation}
where the smallest non-vanishing harmonic is $s_N$ and we assume that also $s_{N+1}$ is not vanishing.  For a generic ideal band, we can expect all $s_n$ to be non-vanishing and $N=1$.
For a flux insertion to map $s_n$ to either $\pm s_n$, $\phi_1, \phi_2$ have to be a multiple of $\pi/N$ and $\pi/(N+1)$. As these numbers are coprime, we conclude $\phi=\pi$ (if $N=1$, $\phi_1, \phi_2$ have to be a multiple of $\pi/N=\pi$ anyway, so in that case, we do not require $s_{N+1}$ to be non-zero.).

Therefore, the only constraint by this flux insertion is that taking
\begin{equation}
    s_n\rightarrow \begin{cases}
        -s_n & \text{ $n$ odd,}\\
        s_n & \text{ else}
    \end{cases}
\end{equation}
leaves the shape of the dispersion as well as its bandwidth invariant.

In the special case where only $s_1,s_2$ are non-zero, we can conclude that the bandwidth only depends on $|s_1|,s_2$. Indeed, as shown in the main text, at fixed $s_1$ the bandwidth of the anyon dispersion is suppressed with increasing $s_2$. On the contrary, a decreasing $s_2$ enhances the bandwidth.

\begin{figure}
    \centering
    \includegraphics[width=0.8\linewidth]{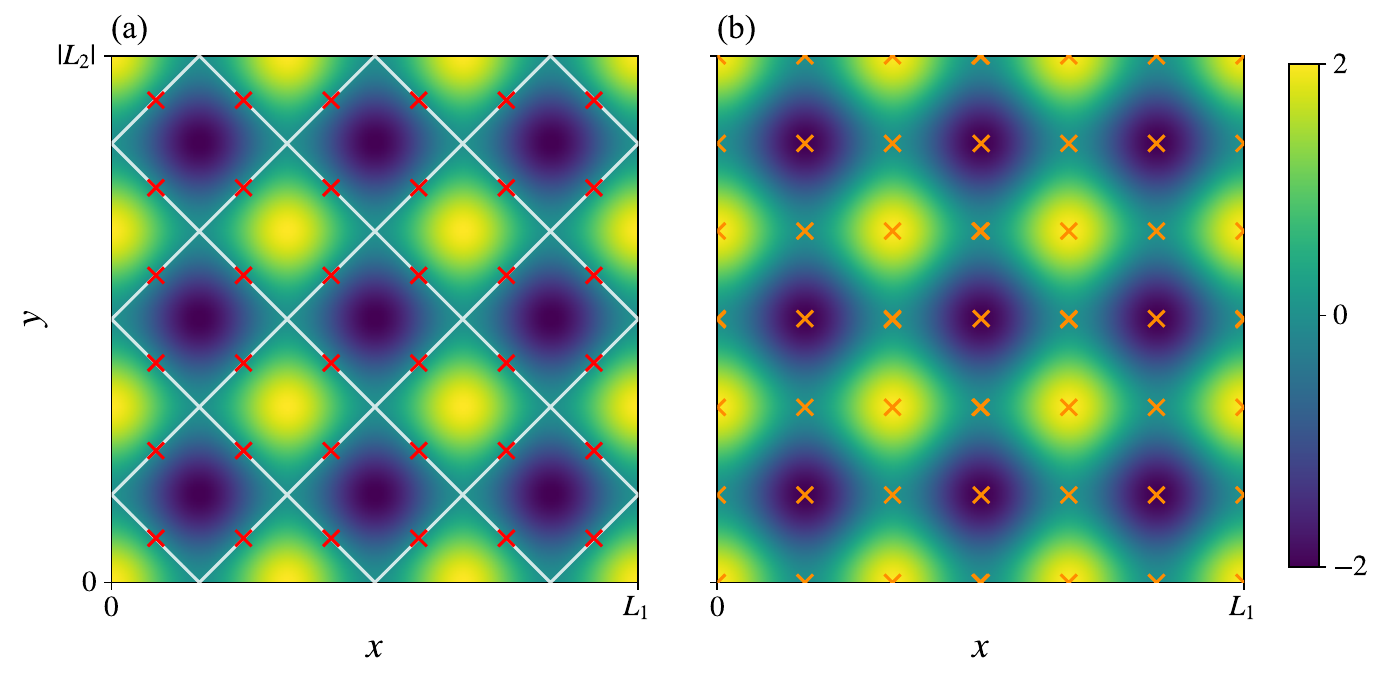}
    \caption{\textbf{Interplay between K\"ahler potential harmonics.} First harmonic of the K\"ahler potential overlaid with the minima of the second harmonic for $s_2>0$ [(a)] and $s_2<0$ [(b)]. For $s_2>0$, the second harmonic minima (red crosses) all coincide with the nodes of the first harmonic (white lines). Hence, as $s_2$ is increased, the influence of the first harmonic on the anyon dispersion gradually decreases moving toward an $s_2$ dominated regime. For $s_2 <0$, some minima of the second harmonic (orange crosses) coincide with the first harmonic minima (dark regions). This deepens the first harmonic minima,  leading to an enhanced dispersion.   }
    \label{fig:kahler_minima}
\end{figure}

This behavior is understood by looking at the minima of the two harmonics.  As shown in Figure \ref{fig:kahler_minima}(a), for $s_2 > 0$,  
the minima of the second harmonic overlaps with the nodes of the first harmonic. Hence, at a fixed $s_1$, increasing $s_2$ reduces the influence of first harmonic leading to a second harmonic dominated dispersion, which as shown in the main text and Appendix \ref{app:infinite_sn_limit} has a negligible dispersion compared to that due to the first harmonic. For $s_2<0$, some minima of the second harmonic overlap with those of the first harmonic, as shown in Figure \ref{fig:kahler_minima}(b). Since the first harmonic minima are reinforced, this leads to a stronger dispersion compared to that at $s_2 = 0$. 

We want to emphasize that this result is special to the rectangular lattice. On e.g.\ a triangular lattice as the one in tMoTe$_2$, the minima (maxima) of the K\"ahler potential for e.g.\ $s_1$ depending on the sign form a triangular (hexagonal) lattice. Flux insertion therefore does not take $s_1\rightarrow -s_1$ and we cannot make any claims about the symmetry of the bandwidth with respect to the sign of $s_1$, consistent with the  results obtained in Ref. \cite{shi2025effectsberrycurvatureideal}.

\section{Exact diagonalization}\label{sup:ED}
In this Appendix, we present additional details of the exact diagonalization (ED) results. To compare and benchmark with Monte-Carlo, we have presented our results for the anyon wavefunctions in Eq.~\eqref{eq:quasihole_momentum_eigenfunction} in the main text. The anyon wavefunctions are generated as exact zero modes of the $V_0$ ($V_1$) Haldane pseudopotentials~\cite{haldane1983} for $m=2$ ($m=3$). Once these eigenstates have been obtained, we can easily evaluate their Coulomb expectation value. Additionally, we can move away from model states and directly consider the exact Coulomb ground states, testing the validity of the model wavefunction approach.

\begin{figure}[t]
    \centering
    \includegraphics[width=1.0\linewidth]{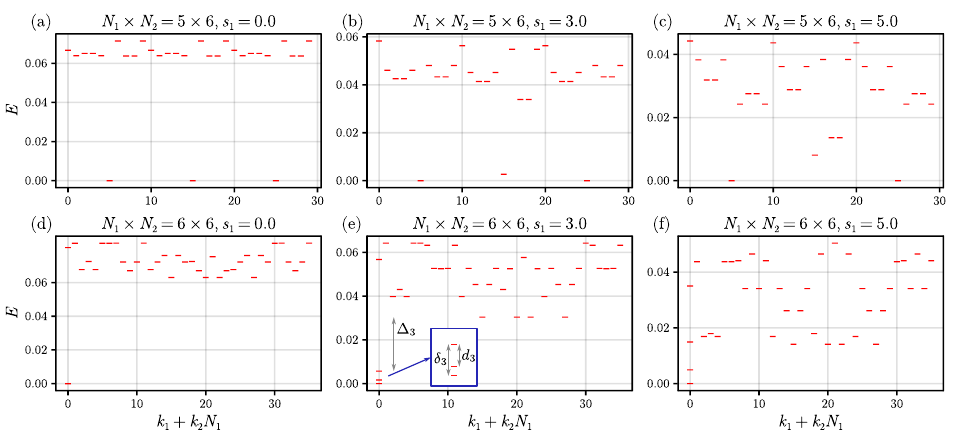}
    \caption{\textbf{Energy spectra.} Energy spectra of the Coulomb interaction on a rectangular torus for $m=3$, $N_1=5$, $N_2=6$ [(a)-(c)] and $N_1=N_2=6$ [(d)-(f)] for $s_1=0,3$ and $5$ from left to right. The energies  are shown in units of $e^2/(\epsilon l)$. For $s_1=0$, we observe an exact three-fold degeneracy of the Laughlin state that is broken at finite $s_1$. For $s_1\sim 5$, the quasi-degeneracy of the Laughlin state is lifted and we expect a phase transition to a symmetry-broken state. We define the indirect gap $\Delta_3$, spread $\delta_3$ and minimal energy difference $d_3$ for the three Laughlin states as depicted in (e).}
    \label{fig:energy_spectra_coulomb}
\end{figure}

\begin{figure}[b]
    \centering
    \includegraphics[width=0.8\linewidth]{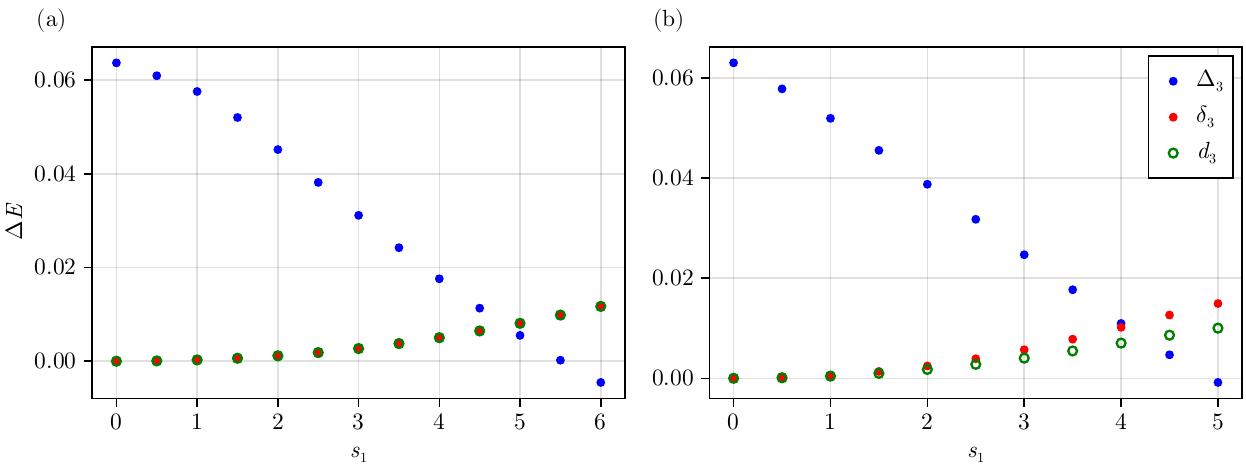}
    \caption{\textbf{FCI gaps.} Indirect gap $\Delta_3$, spread $\delta_3$ and minimal energy difference $d_3$ for the three Laughlin states under Coulomb interaction as defined in Figure~\ref{fig:energy_spectra_coulomb}(e) for $m=3$, $N_1=5$, $N_2=6$ [(a)] and $N_1=N_2=6$ [(b)]. The energies  are shown in units of $e^2/(\epsilon l)$. Depending on the definition of stability for an FCI, commonly $d_3/\Delta_3>1$, $\delta_3/\Delta_3>1$ or $\Delta_3<0$, we observe a break down around $s_1=4-5.5$ for the two system sizes.}
    \label{fig:fci_gaps}
\end{figure}

We begin by analyzing the stability of the FCI for the Laughlin $1/3$ state on systems of size $5\times6$ and $6\times 6$ by analyzing the energy spectra in Figure~\ref{fig:energy_spectra_coulomb} for different values of $s_1$. On both system sizes, we can observe a break-down of the $3$-fold quasidegeneracy around $s_1\sim 5$, for which we show the gap evolution in Figure~\ref{fig:fci_gaps}.

\subsection{Flux threading}\label{app:flux_insertion}
In the main paper, we use flux threading in Figure~1 and Figure~2 to increase the resolution for the quasi-hole dispersion and bandwidth. Threading a flux $\phi_{i}$ in each direction shifts the single-particle momenta as $\mathbf{k}\rightarrow \mathbf{k}+\frac{\phi_i}{2\pi N_i}\mathbf{b}_i$. As the quasiholes have charge $e^*=-e/m$, their many-body momenta are shifted as $k_i\rightarrow k_i+e^*/e \cdot \phi_i$ under flux insertion. For Figure~1 and Figure~2, we insert flux in increments of $\Delta \phi_i = 2\pi/6$ in both directions.

\subsection{Dispersion and bandwidth for the Coulomb ground state}\label{app:coulombED}
\begin{figure}[t]
    \centering
    \includegraphics[width=0.8\linewidth]{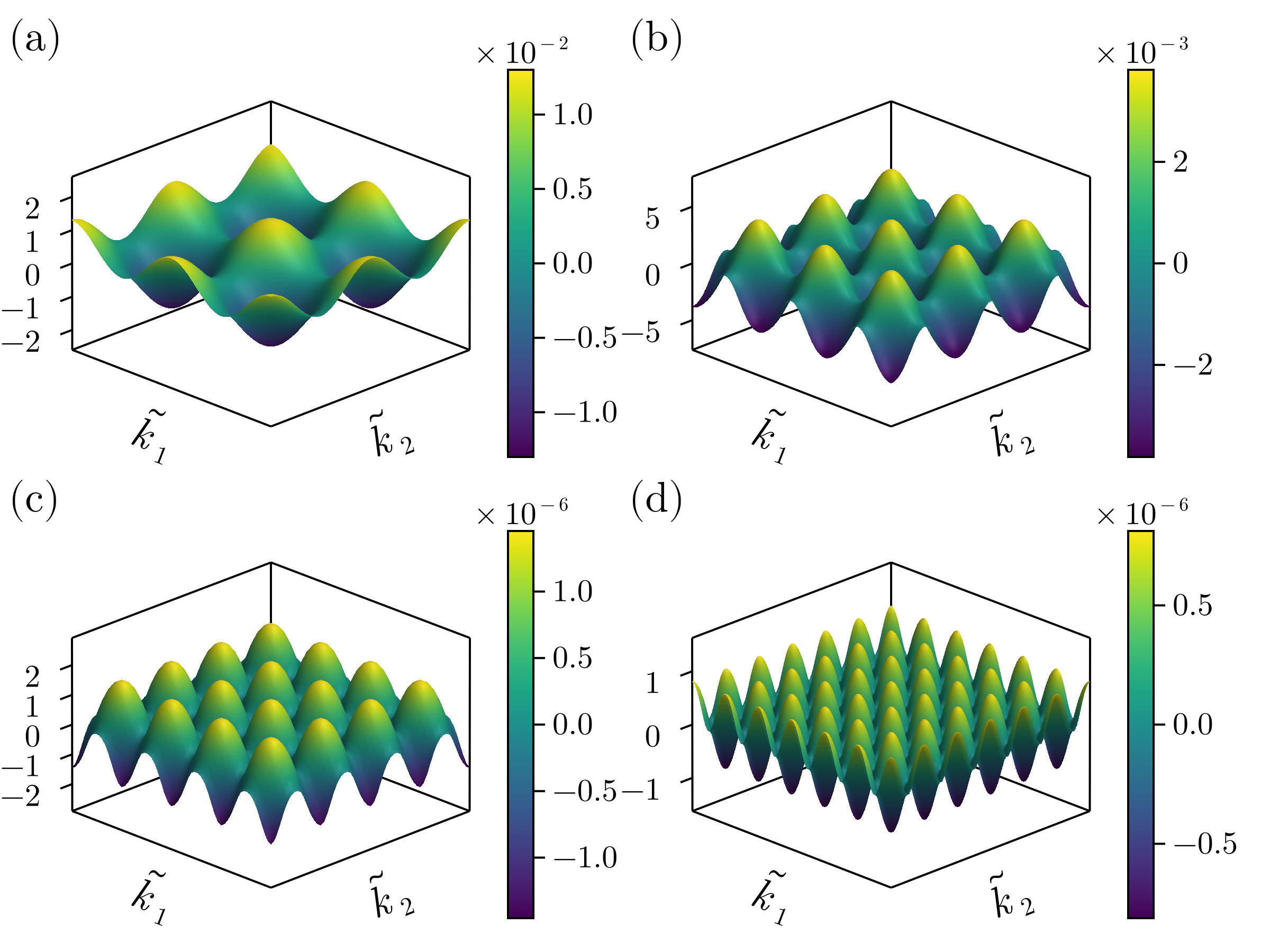}
    \caption{\textbf{Anyon dispersion spectra of the exact ground state.} Coulomb interaction energy of the anyon Coulomb eigenstates on a square torus for the system $m=2$, $N_\phi = 25$ [(a), (c)],  $m = 3$, $N_\phi = 25$  [(b), (d)]. The energies  are shown in units of $e^2/(\epsilon l)$. In (a) \& (b), the energies are plotted for $s_1=1,s_2=0$, showing an $m^2-$fold periodicity in the electronic BZ. In (c)  \& (d),  the energies are plotted for $s_1=0,s_2=1$, showing an enhanced $(2m)^2-$fold periodicity.}
    \label{fig:dispersion_coulomb}
\end{figure}

\begin{figure}[b]
    \centering
    \includegraphics[width=0.8\linewidth]{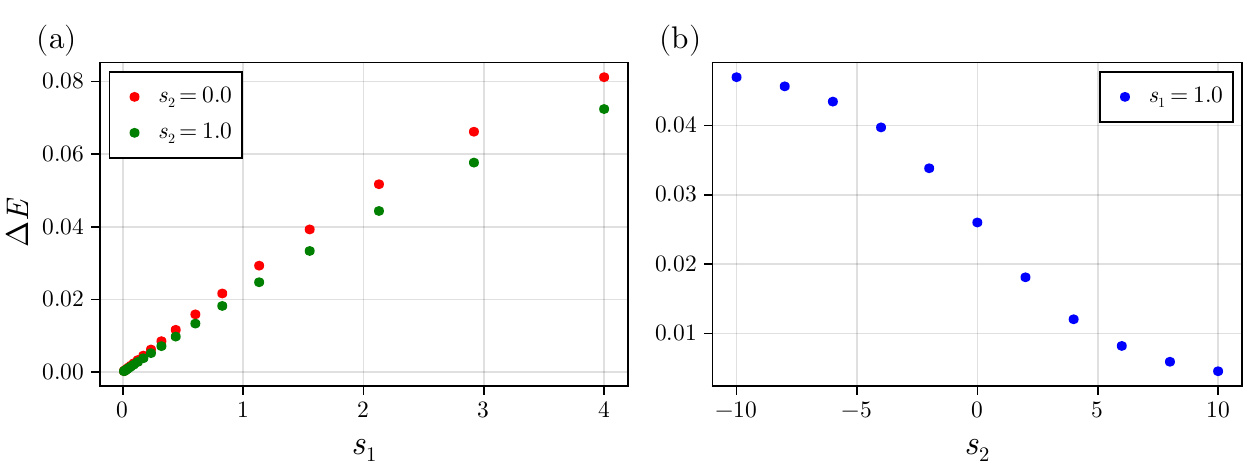}
    \caption{\textbf{Bandwidth evolution with quantum geometry of the exact ground state.} We show bandwidth evolution with quantum geometry for $m=2, N_\phi=25$. (a) At fixed $s_2$, we show the anyon bandwidth as a function of $s_1$. Similarly to the main text, we find the bandwidth to evolve linearly for small $s_1$. A non-zero $s_2$ leads to a smaller bandwidth at the same $s_1$ values. (b) At fixed $s_1$, the bandwidth decreases with positive $s_2$ and increases for negative $s_2$.}   
    \label{fig:bandwidth_coulomb}
\end{figure}

We now present the results for the anyon dispersion and bandwidth for the \emph{exact} Coulomb eigenstates. As we can see in Figure~\ref{fig:dispersion_coulomb} and Figure~\ref{fig:bandwidth_coulomb}, the qualitative results remain the same, only the scale of the anyon dispersion and bandwidth is enhanced compared to the exact anyon states in Eq.~\eqref{eq:quasihole_momentum_eigenfunction}.

\clearpage

\ifSubfilesClassLoaded{
    \bibliography{references}
}{}

\end{document}

\end{document}